\documentstyle[astrobib]{mnv2}
\input psfig

\newcommand{\be}{\begin{equation} }
\newcommand{\ee}{\end{equation} }
\newcommand{\bea}{\begin{eqnarray} }
\newcommand{\eea}{\end{eqnarray} }

\title[Lensing by galaxy halos in clusters of galaxies]
{Lensing by galaxy halos in clusters of galaxies}

\author[Priyamvada Natarajan \& Jean-Paul Kneib]
  {Priyamvada Natarajan$^1$ \& Jean-Paul Kneib$^{2}$ \\
  $^1$Institute of Astronomy, Madingley Road, Cambridge CB3 0HA\\
  $^2$Observatoire Midi-Pyrenees, 14 Av. E.Belin, 31400 Toulouse, France}

\begin{document}
\label{firstpage}
\maketitle

\begin{abstract}
Weak shear maps of the outer regions of clusters have been successfully 
used to map the distribution of mass at large radii from the cluster
center. The typical smoothing lengths employed thus far preclude
the systematic study of the effects of galactic-scale substructure on
the measured weak lensing signal. In this paper, we present two 
methods to infer the possible existence and extent of dark halos
around bright cluster galaxies by quantifying the `local' weak lensing
induced by them. The proposed methods are: direct radial averaging
of the shear field in the vicinity of bright cluster members and 
a maximum-likelihood method to extract fiducial parameters
characterizing galaxy halos. The correlations observed for early-type
galaxies on the Fundamental Plane are used to derive the scaling 
laws with luminosity in the modelling of cluster galaxies. 
We demonstrate using simulations that these
observed local weak-shear effects on galaxy scales within the cluster 
can be used to statistically constrain the mean mass-to-light ratio, 
and fiducial parameters like the halo size, velocity dispersion and
hence mass of cluster galaxies. We compare the two methods and
investigate their relative drawbacks and merits in the context of 
feasibility of application to HST cluster data, whereby we find that 
the prospects are promising for detection on stacking 
a minimum of 20 WFPC2 deep cluster fields.
\end{abstract}

\section{Introduction}

Clusters of galaxies are the most recently assembled structures in the
universe, and the degree of observed substructure in a cluster is the
result of the complex interplay between the underlying cosmological
model (as has been demonstrated by many groups including  
\citeN{bird93}, \citeN{evrard93} and \citeN{west90}) and the physical
processes by which clusters form and evolve. Many clusters
have more than one dynamical component in the velocity 
structure in addition to spatial subclustering (\citeNP{colless95},
\citeNP{kriessler95}, \citeNP{bird93}, \citeNP{west90} and 
\citeNP{fitchett88}). Substructure in the underlying cluster potential
and specifically the sub-clumping of mass on smaller-scales (galactic
scales) within the cluster can be directly mapped via lensing effects.

The observed gravitational lensing of the faint
background population by clusters is increasingly
becoming a promising probe of the detailed mass distribution within
a cluster as well as on larger scales (super-cluster scales). We
expect on theoretical grounds and do observe
local weak shear effects around individual bright galaxies in
clusters over and above the global shearing produced by the `smooth' 
cluster potential. While there is ample evidence from lensing for the
clumping of dark matter on different scales within the cluster, the
spatial extent of dark halos of cluster galaxies are yet to be constrained. 
The issue is of crucial importance as it addresses the key question
of whether the mass to light ratio of galaxies is a function of the 
environment, and if it is indeed significantly different in the high density
regions like cluster cores as opposed to the field. Moreover, it is
the physical processes that operate within  clusters like ram-pressure
stripping, merging and ``harassment'' that imply re-distribution of
mass on smaller scales and their efficiency can be directly probed using 
accurate lensing mass profiles.

Constraining the fundamental parameters such as mass and halo size from
lensing effects for field galaxies was attempted first by
\citeN{tyson84} using plate material, the quality of which precluded
any signal detection. More recently, \citeN{brainerd95} used deep
ground-based imaging and detected the galaxy-galaxy lensing  
signal and hence placed upper limits on the mean mass of an average 
field galaxy. \citeN{griffiths96} used the Medium Deep Survey (MDS)
and HST archival data in a similar manner to extract the
polarization signal. Although the signal is unambiguously detected, 
it is weak, and no strong constraints can yet be put on the mean profile of 
field galaxies, but the prospects are promising for the near future.

On the other hand no such analysis has been pursued in dense regions
like clusters, and very little is known about the lensing effect of
galaxy halos superposed on the lensing effect of a cluster.
\shortciteN{kneib96} have demonstrated the importance of galaxy-scale 
lenses in the mass modeling of the cluster A2218, where the effect of
galaxy-scale components (with a mean mass to light ratio $\sim$ 9 in
the R-band) needs to be included in order to reproduce the observed multiple
images. Mass modeling of several other clusters has also required the
input of smaller-scale mass components to consistently explain the
multiple images as well as the geometry of the arcs, for instance, 
in the case of CL0024 (\citeN{kovner93}, \citeN{smail96}), where the
length of the three images of the cusp arc can only be explained if 
the two nearby bright galaxies contribute mass to the system.
This strongly suggests that the dark matter associated with
individual galaxies is of consequence in accurately mapping 
the mass distribution, and needs to be understood better, particularly
if clusters are to be used as gravitational telescopes to study
background galaxies. 

The observed quantities in cluster lensing studies are the magnitudes 
and shapes of
the background population in the field of the cluster. To reconstruct
the cluster mass distribution there are many 
techniques currently available which allow the inversion of
the distortion map into a relative mass map or an absolute mass map if
(i) multiple arcs are observed (\citeNP{kneib96}) and or (ii) magnification effects are included (\citeNP{broadhurst}). Recent theoretical 
work (\citeNP{kaiser93}, \citeNP{kaiser95a}, \citeNP{peter95a},
\citeNP{caro95a} and \citeNP{squires96}) 
has focused on developing various algorithms to recover
the mass distribution on scales larger than 20-30 
arcsec, which is roughly the smoothing scale employed 
(corresponding to $\sim 100\,{\rm kpc}$ at a redshift of $z\,\sim\,0.2$). These
methods involve locally averaging the shear field produced by the
lensing mass, and cannot be used to probe galaxy-scale
perturbations to the shear field. 

Our aim in this paper is to understand and determine the parameters
that characterize galaxy-scale perturbations within a cluster. 
In order to do so, we delineate 2 regimes:\\
(i) the `strong' regime where the local surface density is close to critical
($\kappa \sim$ 1, where $\kappa$ is the ratio of the local surface
density to the critical surface density) and (ii) the `weak' regime
where the local surface density is small ($\kappa\,<\,1$).
The `strong' regime corresponds to the cores of clusters, and in general
involves only a small fraction (typically 5-20) of the cluster galaxies
whereas the `weak' regime encompasses a larger fraction ($\sim$
50-200). We are restricting our treatment to early-type (E \& S0's)
bright cluster galaxies throughout. 

We compare in this analysis the relative merits of our 2 proposed
methods: a direct method to extract the strength of the averaged local
shear field in the vicinity of bright cluster galaxies by subtracting
the mean large-scale shear field, and a statistical maximum likelihood
method. The former method affords us a physical understanding, helps
establish the importance and the role of the various relevant
parameters and yields a mean mass-to-light ratio; the latter permits
taking the strong lensing regime and the ellipticity of the mass of 
galaxy halos into account correctly. Both approaches are investigated 
in detail in this paper using numerical simulations.

The outline of the paper is as follows. In Section 2, we present the
formalism that takes into account the effect of individual
galaxy-scale perturbations to the global cluster potential.
In Section 3, the direct method to recover these small-scale distortions
is outlined and in Section 4 we present the results of the application
of these techniques to a simulated cluster with substructure.
In Section 5, we examine the constraints that can be obtained on the 
parameter space of models via the proposed maximum-likelihood method. We also 
explore the feasibility criteria for application to cluster data given the
typical uncertainties. The conclusions of this study and the
prospects for application to real data and future work are discussed 
in Section 6. Throughout this paper, we have assumed 
$H_{0} = 50\,$kms$^{-1}$Mpc$^{-1}$, $\Omega = 1$ and $\Lambda = 0$.

\section{Galaxy-Scale Lensing Distortions in Clusters}

\subsection{Analysis of the local distortions}

The mass distribution in a cluster of galaxies can be modeled as 
the linear sum of a global smooth potential (on scales larger than 
20 arcsec) and  perturbing mass distributions which can then be 
associated with individual galaxies (with a scale length less 
than 20 arcsec). 
Formally we write the global potential as:
\be
\phi_{\rm tot} = \phi_{\rm c} + \Sigma_i \,\phi_{\rm p_i},
\ee
where $\phi_{\rm c}$ is the smooth potential of the cluster and 
$\phi_{\rm p_i}$ are the potentials of the perturbers (galaxy halos).
Henceforth, the use of the subscripts c and p refer to quantities
computed for the cluster scale component and the perturbers
respectively. The deflection angle is then given by,
\bea
\theta_S\,=\,\theta_I\,-\,\alpha_I(\theta_I)\ ;
\ \alpha_I\,=\,{{\bmath \nabla}\phi_{\rm c}}\,+\,\Sigma_i \,{{\bmath \nabla}\phi_{\rm p_i}},
\eea
where $\theta_I$ is the angular position of the image and $\theta_S$
the angular position of the source. The amplification matrix at any
given point is,
\be
A^{-1}\,=\,I\,-\,{{\bmath \nabla\nabla} {\phi_{\rm c}}}\,-
\,\Sigma_i \,{{\bmath \nabla\nabla} {\phi_{\rm p_i}}}.
\ee
Defining the generic symmetry matrix,
\begin{displaymath}
J_{2\theta}\,=\, \left(\begin{array}{lr}
\cos {2\theta}&\sin {2\theta}\\
\sin {2\theta}&-\cos {2\theta}\\
\end{array}\right)
\end{displaymath}
we decompose the amplification matrix as a linear sum:
\bea
A^{-1}\,=\,(1\,-\,\kappa_{\rm c}\,-\,\Sigma_i \kappa_{\rm p})\,I
- \gamma_{\rm c}J_{2\theta_{\rm  c}}
- \Sigma_i \,\gamma_{\rm p_i}J_{2\theta_{\rm  p_i}},
\eea
where $\kappa$ is the magnification and $\gamma$ the shear. 
In this framework, the shear $\gamma$ is taken to be a complex number
and is used to define the quantity $\overline{g}$ as follows:
\bea
\overline{g_{pot}} = {\overline{\gamma} \over 1-\kappa} =
{{\overline\gamma_c} + \Sigma_i \,{\overline\gamma_{p_i}}
 \over 1-\kappa_c -\Sigma_i
 \,\kappa_{p_i}},\,\,{\overline{\tau_{pot}}}\,=\,
 { 2\overline{g_{pot}}
 \over 1 - \overline{g_{pot}}^*\overline{g_{pot}}}
\eea
which simplifies in the frame of the perturber $j$ to (neglecting
effect of perturber $i$ if $i \neq j$):
\bea
{\overline g_{pot}}|_j} = 
{ {{\overline \gamma_c} +{\overline \gamma_{p_j}} \over {1-\kappa_c -\kappa_{p_j}}},
\eea
where $\overline g_{pot}|_j$ is the total complex shear induced by
the smooth cluster potential and the potentials of the perturbers.
Restricting our analysis to the weak regime, and thereby retaining
only the first order terms from the lensing equation for the shape
parameters (see \citeNP{kneib96}) we have:
\be
{\overline \tau_I}= {\overline \tau_S}+{\overline \tau_{pot}},
\ee
where ${\overline \tau_I}$ is the distortion of the image, ${\overline
\tau_S}$ the intrinsic shape of the source, ${\overline \tau_{\rm pot}}$
is the distortion induced by the lensing potentials or explicitly 
in terms of $\overline g_{pot}$ in the frame of perturber $j$:
\be
{\overline g_I}= {\overline g_S}+{\overline g_{pot}}|_j
= {\overline g_S} +
{ {\overline \gamma_c} \over 1-\kappa_c -\kappa_{p_j}} +
{ {\overline \gamma_{p_j}} \over 1-\kappa_c - \kappa_{p_j}}.
\ee
In the local frame of reference of the perturbers, the mean
value of the quantity ${\overline g_I}$ and its dispersion can be 
computed in circular annuli (of radius $r$ from the perturber center)
{\underline{strictly in the weak-regime}},
assuming a constant value $\gamma_c e^{i\theta_{c0}}$ for the smooth 
cluster component over the area of integration (see Figure 1 for the
schematic diagram). 
\begin{figure*}
\psfig{figure=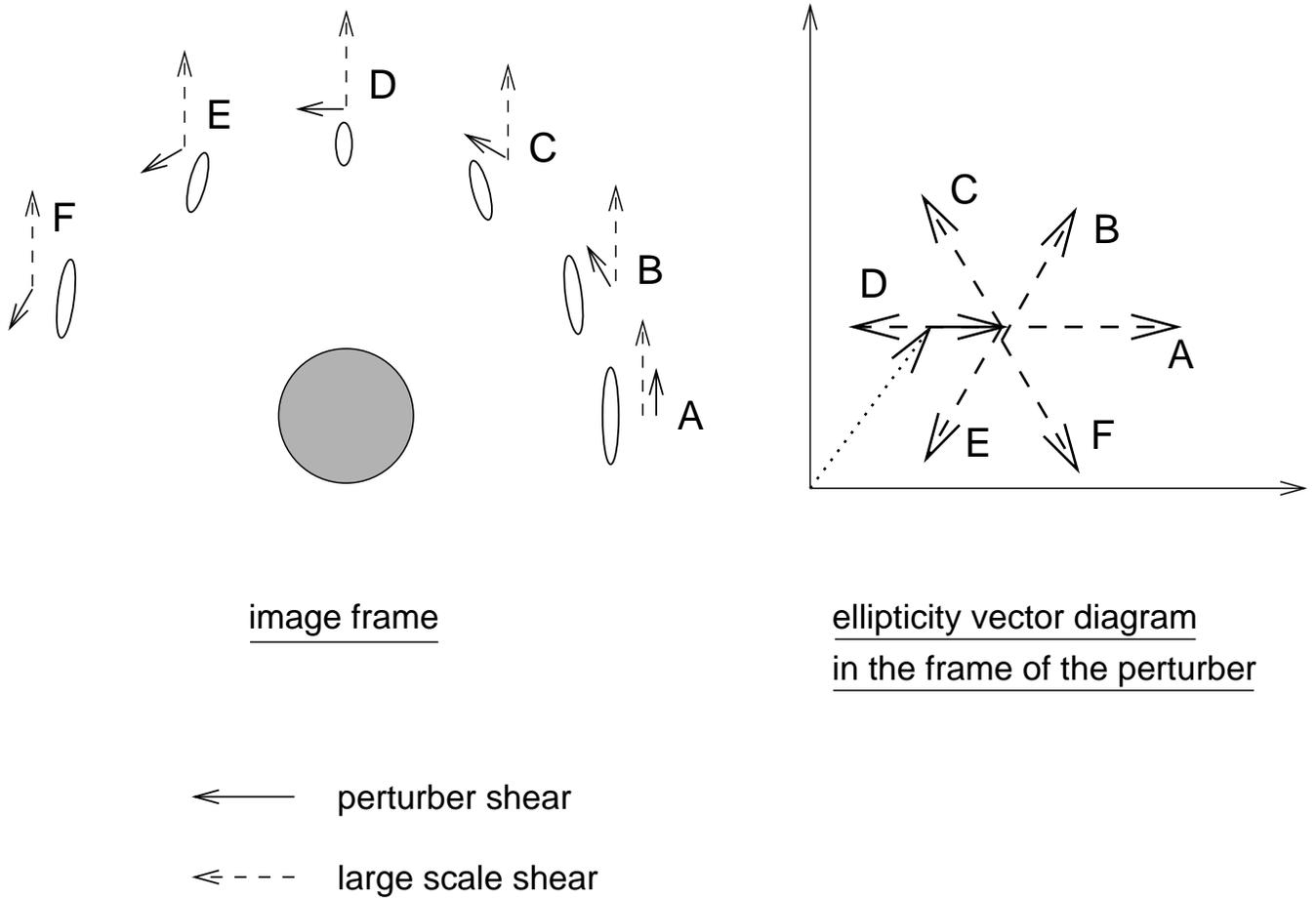,width=1.0\textwidth}
\caption{Local frame of reference of the perturber: The vector diagram
illustrating the choice of coordinate system. The total shear is
decomposed into a large-scale component due to the smooth cluster
and a small-scale one due to the perturbing galaxy. In the frame of
the perturber, the averaging procedure allows efficient subtraction of
the large-scale component as shown in the right panel, enabling the
extraction of the shear component induced in the background galaxies
only by the perturber as shown in the left panel. The background
galaxies (shown in the left panel of this figure) are assumed to have
the same intrinsic ellipticity for simplicity, therefore, we plot only
the induced components.}
\end{figure*}

\begin{figure*}
\psfig{figure=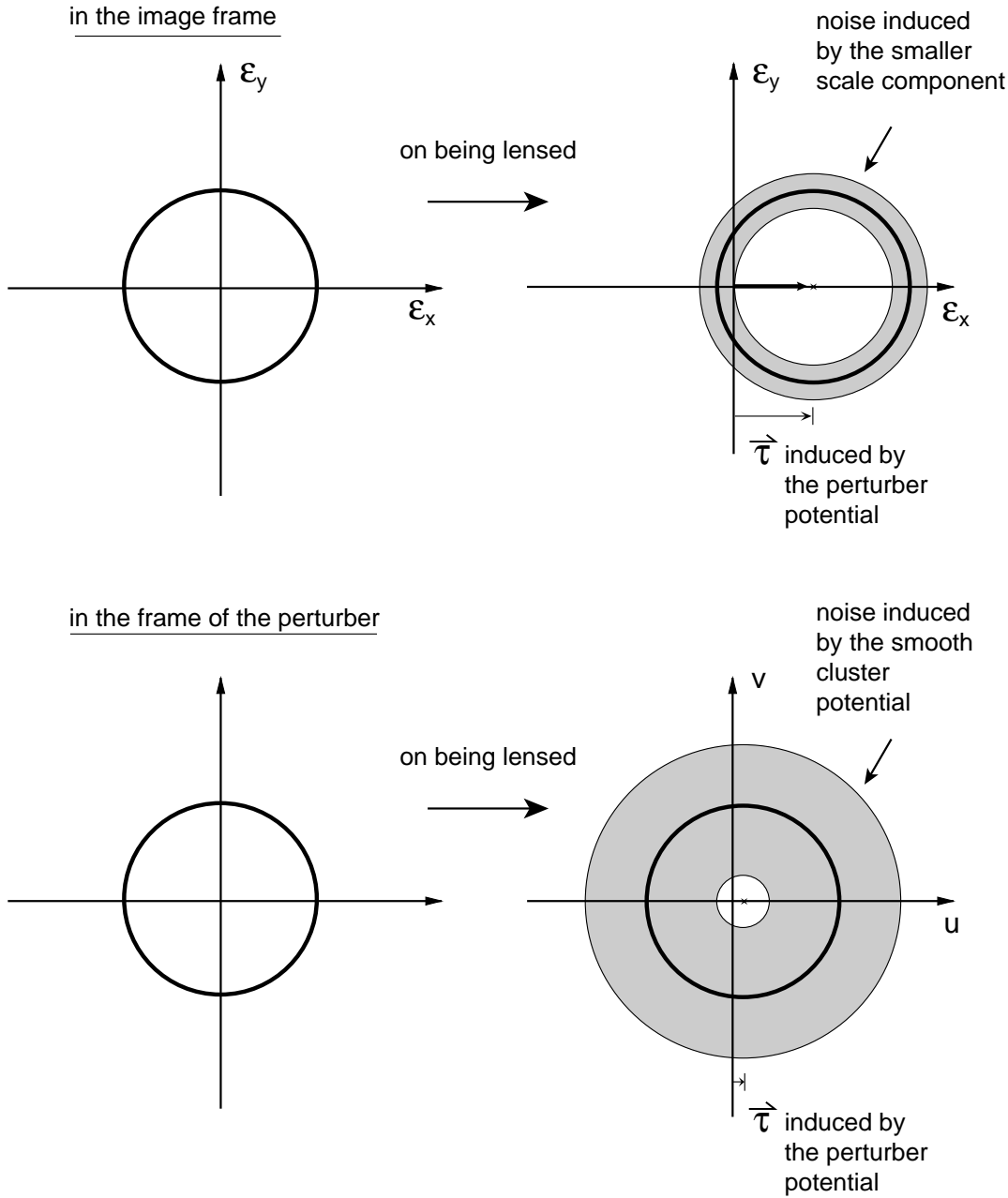,width=0.8\textwidth}
\caption{A schematic representation of the effect of the cluster
on the intrinsic ellipticity distribution of background sources as viewed from
the two different frames of reference. In the top panel, as viewed in
the image frame - the effect of the cluster is to cause a coherent
displacement $\bmath{\tau}$ and the presence of perturbers merely adds
small-scale noise to the observed ellipticity distribution. In the bottom panel,
as viewed in the perturbers frame - here the perturber component
causes a small displacement $\bmath{\tau}$ and the cluster component
induces the additional noise.}
\end{figure*}

The result of the integration does depend on the choice of coordinate system. 
In cartesian coordinates (averaging out the contribution of the perturbers):
\bea
\nonumber \left<{\overline g_I}\right>_{xy} &=& \left<{\overline g_S}\right> + 
\left<{\gamma_c e^{i\theta_{c0}} \over 1-\kappa_c -\kappa_{p_j}}\right> +
\left<{{\overline \gamma_{p_j}} \over 1-\kappa_c - \kappa_{p_j}}\right>,
 \\ \nonumber
&=& {\gamma_c e^{i\theta_{c0}}} \left<{1 \over 1-\kappa_c -\kappa_{p_j}}\right>
\equiv {\overline g_c},\\
\eea
\be
\sigma^2_{\overline g_I} = {\sigma^2_{\overline g_S} \over  2}
+ {\sigma^2_{\overline g_{p_j}} \over 2},
\ee
where 
\bea
\sigma^2_{g_I}\,\approx\,{\sigma^2_{p(\tau_S)}\over 2 N_{bg} } 
+ { \sigma^2_{\overline {g}_{p_j}} \over 2 N_{bg} }\,\approx\,
{\sigma^2_{p(\tau_S)}\over 2 N_{bg}}
\eea
${\sigma^2_{p(\tau_S)}}$ being
the width of the intrinsic ellipticity distribution of the sources,
$N_{bg}$ the number of background galaxies averaged over and
$\sigma^2_{\overline {g}_{p_j}}$ the dispersion due to perturber effects
which should be smaller than the width of intrinsic ellipticity distribution. 
In the polar $uv$ coordinates, on averaging out the smooth part:
\bea
\nonumber \left<{\overline g_I}\right>_{uv} &=& \left<{\overline g_S}\right> +
\left<{{\overline \gamma_c} \over 1-\kappa_c - \kappa_{p_j}}\right> +
\left<{\gamma_{p_j}\over 1-\kappa_c -\kappa_{p_j}}\right>,
 \\ \nonumber
&=& {\gamma_{p_j}}\left<{ 1 \over {1-\kappa_c -\kappa_{p_j}}}\right>\equiv g_{p_j},\\
\eea
\bea
\left(\sigma^2_{\overline{g_I}}\right)_{uv} = {{\sigma^2_{\overline g_S}} \over 2}
+ {{\sigma^2_{\overline g_c}} \over 2},
\eea
where 
\bea
\sigma^2_{g_I}\,\approx\,{\sigma^2_{p(\tau_S)}\over 2 N_{bg} } +
{\sigma^2_{\overline{g}_{c}} \over 2 N_{bg} }.
\eea
 From these equations, we clearly see the two effects of the
contribution of the smooth cluster component: it
boosts the shear induced by the perturber due to the
($\kappa_c+\kappa_{p_j}$) term in the denominator, which becomes non-negligible 
in the cluster center, 
and it simultaneously dilutes the regular galaxy-galaxy lensing signal
due to the ${\sigma^2_{\overline g_c} / 2}$ term (equation 11) in the dispersion of
the polarization measure. However, one can in principle optimize
the noise in the polarization by `subtracting' the measured cluster signal
and averaging it in polar coordinates:
\be
\left<{\overline g_I}-{\overline g_c}\right>_{uv} = 
\left<{\gamma_{p_j}\over 1-\kappa_c -\kappa_{p_j}}\right>,
\ee
which gives the same mean value as in equation (11) but with a reduced
dispersion:
\be
\left(\sigma^2_{\overline g_I-\overline g_c}\right)_{uv} 
= {\sigma^2_{\overline g_S} \over 2},
\ee
where 
\bea
\sigma_{g_S}^2\,\approx\, {\sigma^2_{p(\tau_S)}\over 2 N_{bg}}.
\eea
This subtraction of the larger-scale component reduces the noise in
the polarization measure, by about
a factor of two; when $\sigma^2_{\overline g_S}\sim \sigma^2_{\overline
g_c}$, which is the case in cluster cores. Note that in subsequent
sections of the paper, we plot the averaged components of ${\overline{\tau}}$
(the quantity measurable from lensing observations) computed in the
(uv) frame. We reiterate here that the calculations above assume that
the cluster component is constant over the area of integration (a
reasonable assumption if we limit our analysis to small radii around
the centers of perturbers). 
These results can be easily extended to the case when the cluster 
component is linear (in $x$ and $y$) over the area of integration, 
the likely case outside the core region. This direct averaging prescription for
extracting the distortions induced by the possible presence of dark halos
around cluster galaxies, by construction, does not require precise knowledge 
of the center of the cluster potential well.

\subsection{Quantifying the lensing distortion}

To quantify the lensing distortion induced by the individual
galaxy-scale components using a minimal number of parameters to
characterize cluster galaxy halos, we model the density profile as a 
linear superposition of two
pseudo-isothermal elliptical components (PIEMD models derived by \citeNP{kassiola93}): 
\bea
\Sigma(R)\,=\,{\Sigma_0 r_0  \over {1 - r_0/r_t}}
({1 \over \sqrt{r_0^2+R^2}}\,-\,{1 \over \sqrt{r_t^2+R^2}}),
\eea
with a model core-radius $r_0$ and a truncation radius $r_t\,\gg\, r_0$.
The useful feature of this model, 
is the ability to reproduce a large range of mass
distributions by varying only the ratio $\eta$: defined as
$\eta=r_t/r_0$. It also provides the following simple relation between
the truncation radius and the effective radius $R_{\rm e}$, $r_t\sim
(4/3) R_{\rm e}$.
Furthermore, this apparently circular model can be easily generalized 
to the elliptical case by re-defining the radial coordinate $R$ as follows:
\bea
R^2\,=\,({x^2 \over  (1+\epsilon)^2}\,+\,{y^2 \over (1-\epsilon)^2})\,;
\ \ \epsilon= {a-b \over a+b},
\eea
The mass enclosed within radius $R$ for the model is given by:
\be
M(R)={2\pi\Sigma_0 r_0 \over {1-{{r_0} \over {r_t}}}}
[\,\sqrt{r_0^2+R^2}\,-\,\sqrt{r_t^2+R^2}\,+\,(r_t-r_0)\,],
\ee
and the total mass, which is finite, is:
\be
{M_{\infty}}\,=\,{2 \pi {\Sigma_0} {r_0} {r_t}}.
\ee
Calculating $\kappa$, $\gamma$ and $g$, we have,
\bea
\kappa(R)\,=\,{\kappa_0}\,{{r_0} \over {(1 - {r_0/r_t})}}\,
({1 \over {\sqrt{({r_0^2}+{R^2})}}}\,-\,{1
\over {\sqrt{({r_t^2}+{R^2})}}})\,\,\,,
\eea
\bea
2\kappa_0\,=\,\Sigma_0\,{4\pi G \over c^2}\,{D_{\rm ls}D_{\rm ol}
  \over D_{\rm os}},
\eea
where $D_{\rm ls}$, $D_{\rm os}$ and $D_{\rm ol}$ are respectively
the lens-source, observer-source and observer-lens angular diameter distances.
To obtain $g(R)$, knowing the magnification $\kappa(R)$, we solve
Laplace's equation for the projected potential $\phi_{\rm 2D}$,
evaluate the components of the amplification matrix and then proceed
to solve directly for $\gamma(R)$, $g(R)$ and $\tau(R)$. 
\bea
\phi_{2D}\,&=&\, \nonumber 2{\kappa_0}[\sqrt{r_0^2+R^2}\,-\,\sqrt{r_t^2+R^2}\,+
(r_0-r_t) \ln R\, \\ \nonumber \,\, 
&-&
\,r_0\ln\,[r_0^2+r_0\sqrt{r_0^2+R^2}]\,+
\,r_t\ln\,[r_t^2+r_t\sqrt{r_t^2+R^2}] ].\\
\eea
To first approximation,
\bea
\tau(R)\,\approx\,\gamma(R)\,
&=&\,\nonumber 
\kappa_0[\,-{1 \over \sqrt{R^2 + r_0^2}}\,
        +\,{2 \over R^2}(\sqrt{R^2 + r_0^2}-r_0)\,\\
\nonumber 
&+&\,{1 \over {\sqrt{R^2 + r_t^2}}}\,-\,
{2 \over R^2}(\sqrt{R^2 + r_t^2} - r_t)\,].\\
\eea
Scaling this relation by $r_t$ gives for $r_0<R<r_t$:
\be
\gamma(R/r_t)\propto {\Sigma_0 \over \eta-1} {{r_t} \over
  R}\,\sim\,{\sigma^2 \over R},
\ee
where $\sigma$ is the velocity dispersion and for $r_0<r_t<R$:
\be
\gamma(R/r_t)\propto {\Sigma_0\over\eta} {{r_t}^2 \over
  R^2}\,\sim\,{{M_{\rm tot}}
  \over {R^2}},
\ee
where ${M_{\rm tot}}$ is the total mass. In the limit that $R\,\gg\,r_t$, we have, 
\bea 
\gamma(R)\,=\,{{3 \kappa_{0}}  \over {2
{R^3}}}\,[{r_{0}^2}\,-\,{r_{t}^2}]\,+\,{{2 {\kappa_0}} 
\over {R^2}} [{{r_t}\,-\,{r_0}}],
\eea
and as ${R\,\to\,\infty}$, $\gamma(R)\,\to\,0$, $g(R)\,\to\,0$ and
$\tau(R)\,\to\,0$ as expected.
\begin{figure}
\psfig{figure=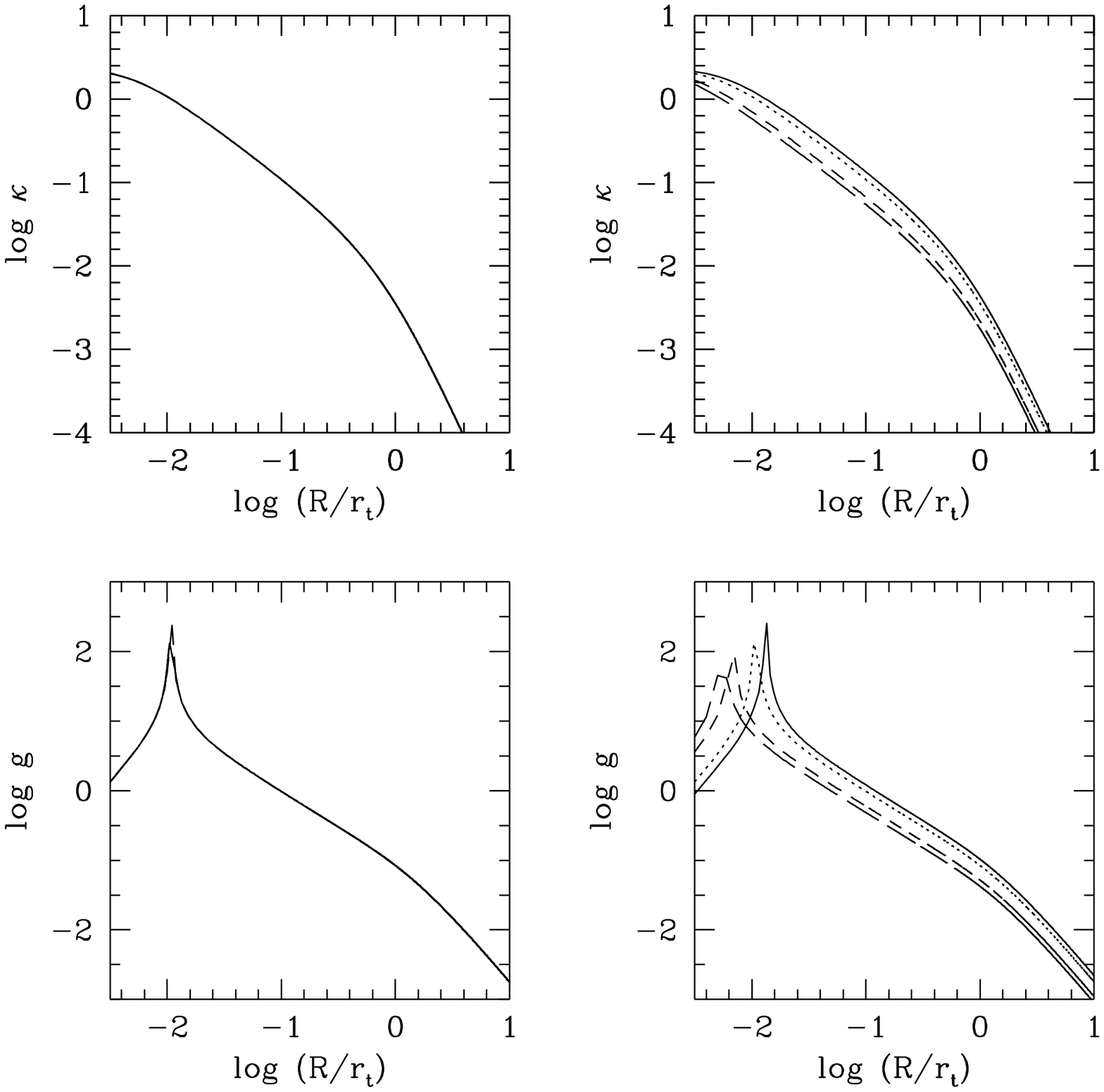,width=0.5\textwidth}
\caption{The effect of the assumed scaling relations are examined in a plot of the
magnification log $\kappa$ vs. ${R/r_t}$ and the shear $g$
vs. ${R/r_t}$ for various values 
of $(L/{L^{*}})$: 0.5, 1.0, 5.0 and 10.0. The curves on the left panel 
are for $\alpha=0.5$ and on the right panel for $\alpha=0.8$, (i) solid
curves - $(L/{L^{*}})$ = 0.5, (ii) dotted curves - $(L/{L^{*}})$ = 1.0,
(iii) short-dashed curves - $(L/{L^{*}})$ = 5.0, (iv) long-dashed curves
- $(L/{L^{*}})$ = 10. The magnification is normalized so that at r = 2
$r_0$, $\kappa$ = 1; the difference in the slope of $\kappa$ above and
below log r/rt = 0 can be clearly seen for both sets of scaling
laws. Note that a spike appears in the plots of log $g$ vs. log
$(R/r_t)$ at the radius where the mean enclosed surface density is
approximately equal to the critical surface density. For the mass
models studied here (cuspy with small core-radii) the surface mass
density has a large central value and hence a spike appears on a scale
that is roughly comparable to the core-radius.}
\end{figure}

\section{Recovering galaxy-scale perturbations}

In this section, we study the influence of the various parameters
using the direct averaging procedure on the synthetic data obtained
from simulations. 
The numerical simulations involve modeling of the global cluster
potential, the individual perturbing cluster galaxies and calculating
their combined lensing effects on a catalog of
faint galaxies. We compute the
mapping between the source and image plane and hence solve the lensing
equation, using the lens tool
utility developed by \citeN{kneib93b}, which accounts consistently for
the displacement and distortion of images both in the strong and weak lensing regimes.

\subsection{Modeling the cluster galaxies}

\subsubsection{Spatial and Luminosity distribution}

A catalog of cluster galaxies was generated at random with the following
characteristics. The luminosities were drawn from a 
standard Schechter function with ${L_*}\,=\,3.{10^{10}}L\odot$ and 
$\alpha=-1.25$. The positions were assigned  
consistent with the number density $\nu(r)$ of a modified Hubble law profile,
\bea
\nu(r)\,=\,{\nu_0  \over {(1+{r^2 \over r_0^2})}^{1.5}},
\eea
with a core radius $r_0=250\,kpc$, as well as a more generic
`core-less' profile of the form:
\bea
\nu(r)\,=\,{\nu_0  \over {{{r \over r_s}^\alpha}(1+{r^2 \over r_s^2})}^{2-\alpha}},
\eea
with a scale-radius ${r_s}=200\,kpc$ and $\alpha\,=\,0.1$ which was
found to be a good-fit to the galaxy data of the moderate redshift
lensing cluster A2218 by \citeN{natarajan96c}. We find however, that
the results for the predicted shear from the simulations is
independent of this choice. 
\subsubsection{Scaling laws}

The individual galaxies are then parameterized by the mass model of Section
2.2, using in addition, the following scalings with luminosity (see
\citeN{brainerd95} for an analogous treatment) for the central velocity dispersion ${\sigma_0}$,
the truncation radius $r_t$ and the core radius $r_0$:
\bea
{\sigma_0}\,=\,{\sigma_{0*}}({L \over L_*})^{1 \over 4}; \\
{r_0}\,=\,{r_{0*}}{({L \over L_*}) ^{1 \over 2}}; \\
{r_t}\,=\,{r_{t*}}{({L \over L_*})^{\alpha}}.
\eea
These imply the following scaling for the $r_t/r_0$ ratio $\eta$:
\be
{\eta}\,=\,{r_t\over r_0}={{r_{t*}} \over {r_{0*}}} 
({L \over L_*})^{\alpha-1/2}.
\ee
The total mass $M$ then scales with the luminosity as:
\be
\,\,M\,=\,{2 \pi {\Sigma_0} {r_0} {r_t}}\,=\,{9\over 2G}(\sigma_0)^2 r_t= 
{9\over 2G}{{\sigma_{0*}}^2}{r_{t*}}({L \over L_*})^{{1 \over 2}+\alpha},
\ee
where $\alpha$ tunes the size of the galaxy halo, and the
mass-to-light ratio $\Upsilon$ is given by:
\bea
{\Upsilon}\,= 12 \left( {\sigma_{0*}\over 240\,km/s}\right)^2
   \left( {r_{t*}\over 30\,kpc} \right)
   \left( {L\over L_*} \right )^{\alpha-1/2}
\eea
Therefore, for  $\alpha$ = 0.5 the assumed galaxy model has constant
$\Upsilon$  for each galaxy; if $\alpha>$ 0.5 ($\alpha<$ 0.5) then
brighter galaxies have a larger (smaller) halos than the fainter
ones. 

The physical motivation for exploring these scaling laws arises from
trying to understand the observed empirical correlations for early-type (E \&
S0) galaxies in the fundamental plane (FP). The following tight relation
between the effective radius $R_e$, the central velocity dispersion
${\sigma_0}$ and the mean surface brightness within $R_e$ is found for
cluster galaxies (\citeN{inger96}, \citeN{djorgovski87} and 
\citeN{dressler87}):
\begin{eqnarray}
\log {R_e}\,=\,1.24\,\log {{\sigma_0}}\,-\,0.82\,\log {{\left<{I}\right>}_e}
+ cste
\end{eqnarray}
One of the important consequences of this relation is the fact that it
necessarily implies that the mass-to-light ratio is a weak function of
the luminosity, typically $\Upsilon\,\sim\,{L^{0.3}}$ (\citeNP{inger96}).
In terms of our scaling scaling laws, this implies $\alpha=0.8$.
Henceforth, in this analysis we explore both the scaling relations, 
for $\alpha\,=\,0.5$; the constant mass-to-light ratio case, 
and $\alpha\,=\,0.8$; corresponding to the mass-to-light ratio being 
proportional to ${L^{0.3}}$ - consistent with the observed FP.
In Figure 3, we plot the scaling relations for
various values of $({L/L_{*}})$, ranging
from 0.5 $\to$ 10.0 for $\alpha$ = 0.5 and $\alpha$ = 0.8. 
\begin{figure}
\psfig{figure=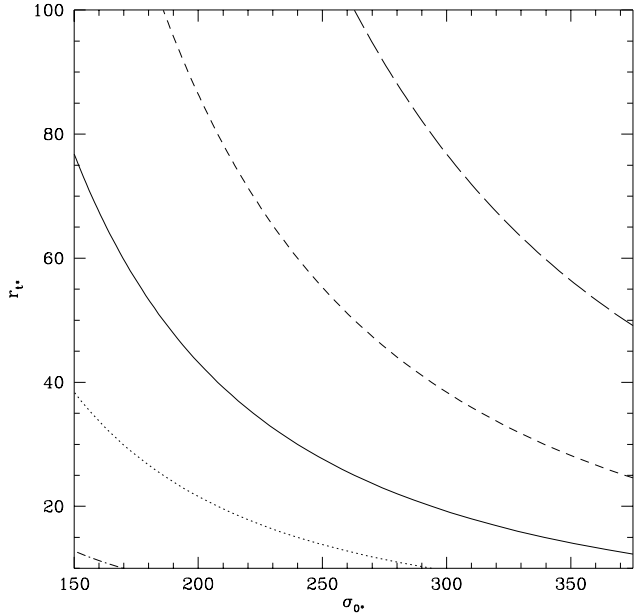,width=0.5\textwidth}
\caption{The constant mass-to-light ratio curves are plotted in the
(${\sigma_{0*}}$,${r_{t*}}$) plane for an ${L_{*}}$ galaxy with $\eta$ =
200: (i) dot-dashed curve - $\Upsilon$ = 4, (ii) dotted curve - 
$\Upsilon$ = 6, (iii) solid curve - $\Upsilon$ = 12, (iv) 
short-dashed curve - $\Upsilon$ = 24 and (v) long-dashed curve - 
$\Upsilon$ = 48.}
\end{figure}
Additionally, for the constant mass-to-light ratio case, we also plot
the iso-$\Upsilon$ curves in terms of the fiducial $\sigma_{0}^{*}$
and $r_{t}^{*}$ in Figure 4. The scaling laws are calibrated by defining an
$L_*$ (in the R-band) elliptical galaxy to have ${r_{0*}}\,=\,$0.15 kpc,
${r_{t*}}\,=\,$30.0 kpc and a fiducial ${\sigma_{0*}}$, then chosen to
assign the different mass-to-light ratios, [${\sigma_{0*}}\,=\,$100,
140, 170, 240, 340, 480 km $s^{-1}$ corresponding to $\Upsilon\,=\,$2,
4, 6, 12, 24, 48 respectively].

\subsection{Modeling the background galaxies}

\subsubsection{Luminosity distribution}

The magnitude and hence the luminosity for the background population
was generated consistent with the number count distribution measured
from faint field galaxy surveys like the MDS 
as reported in \citeN{glazebrook95}, as well as the more recent
results of the number-magnitude relations obtained from the Hubble
Deep Field data (\citeNP{abraham96}). The slope of the number count
distribution used was 0.33 over the magnitude range $m_R = 18 - 26$. 
This power law for the number counts implies a surface number density
that is roughly 90 galaxies per square arcminutes in the given
magnitude range (see \citeN{smail95f}), which over the area of the
simulation frame [8 arcmin X 8 arcmin] corresponds to having $\sim$
5000 background galaxies.
\subsubsection{Redshift distribution}

The background galaxy population of sources was also generated, 
consistent with the measured 
redshift, magnitude and luminosity distributions (MODEL Z2 below) from
high-redshift surveys like the APM and CFRS
(\citeNP{efsta91a} and \citeNP{cfrs195} respectively).
For the normalized redshift distribution at a given magnitude
$m$ (in the R-band) we used the following fiducial forms:
\subsubsection*{{\bf MODEL Z1}:}
\bea
N(z,m)\,=\,{N_0}{\delta(z-2)},
\eea
corresponding to the simple case of placing all the sources at 
$z\,=\,$2. 
\subsubsection*{{\bf MODEL Z2}:}
\bea
N(z,m)\,=\,{{\beta\,({{z^2} \over {z_0^2}})\,
\exp(-({z \over {z_0}})^{\beta})}  \over 
{\Gamma({3 \over \beta})\,{{z_0}}}};
\eea
where $\beta\,=\,$1.5 and
\bea
z_0\,=\,0.7\,[\,{z_{\rm median}}\,+\,{{d{z_{\rm median}}} \over
{dm_R}}{(m_R\,-\,{m_{R0}})}\,],
\eea
${z_{\rm median}}$ being the median redshift, $dz_{\rm median}/ dm_R$
the change in median redshift with R magnitude $m_R$.
We use for our simulations $m_{R0}=$22.0, $dz_{\rm median}/dm_R$=0.1
and $z_{\rm median}$= 0.58 (see \citeN{brainerd95} and \citeN{kneib96}). 

\subsubsection{Ellipticity distribution}

Analysis of deep surveys such as the MDS 
fields (\citeNP{griffiths94}) shows that the ellipticity distribution 
of sources is a strong function of the sizes of individual galaxies as
well as their magnitude (\citeNP{kneib96}). For
the purposes of our simulations, since we assume `perfect seeing', we 
ignore these effects and the ellipticities are assigned in concordance
with an ellipticity distribution $p(\tau_S)$ derived from fits to the MDS data
(\citeNP{ebbels96}) of the form,
\be
p(\tau_S)\,=\,\tau_S\,\,\exp(-({\tau_S \over
\delta})^{\alpha});\,\,\,\alpha\,=\,1.15,\,\,\delta\,=\,0.25.
\ee

\section{\bf Analysis of the Simulations}

We use the above as input distributions to simulate the background
galaxies and bright cluster galaxies in 
addition to a model for the cluster-scale mass
distribution. Analogous to the mass model constructed for the 
cluster Abell 2218 (\citeN{kneib96}), we set up an elliptical mass 
distribution for the central clump with a velocity dispersion of 
1100 km s$^{-1}$ placed at a redshift $z = 0.175$. The main clump was
modelled using a PIEMD profile
(as in equation (14)) with an ellipticity $\epsilon\,=\,0.3$,
core radius 70 kpc and a truncation radius 700 kpc; therefore the
surface mass density of the clump falls off as 
$r^{-3}$ for $r\,\gg\,{r_{\rm cut}}$. 

The lens equation was then
solved for the specified configurations of
sources and lenses set-up as above and the corresponding image frames 
were generated. The averaged components of the shear binned in
circular annuli centred on the perturbing galaxies was evaluated in 
their respective local (u,v) frames. An important check on the entire
recovery procedure arises from the fact that by construction (choice
of the (u,v) coordinate system) the mean value of the v-component of
the shear $<\tau_v>$ is required to vanish. 

In the following sub-sections, we explore the dependence of the 
strength of the detected signal on the various input parameters.  
\begin{figure}
\psfig{figure=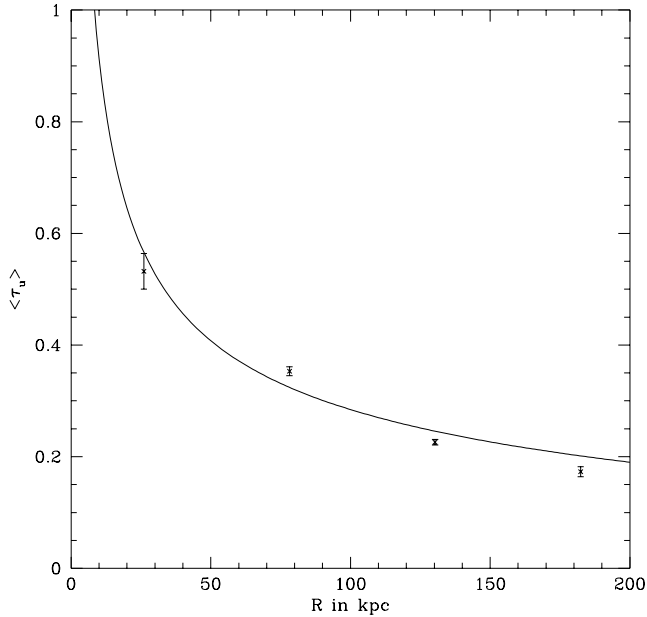,width=0.5\textwidth}
\caption{Demonstrating the robustness of the signal extraction by
  comparing the analytic prediction with the measured radially
  averaged shear from the simulation. The signal was extracted 
from a simulation run of a PIEMD model with ${r_{t*}}$ = 30 kpc, 
${r_{0*}}$ = 0.15 kpc and velocity dispersion of 480 kms$^{-1}$: 
solid curve - estimate from the analytic formula and overplotted are the
measured values of the averaged shear.}
\end{figure}
First of all, Figure 5 demonstrates the good agreement between the 
analytic formula for the shear derived at a given radial distance R
produced by a PIEMD model as computed in Section 2.2 and the averaged value
extracted from the simulation on solving the lensing equation exactly
for the redshift distribution of MODEL Z1. In all subsequent plots
(Figures 6, 7, 8, 9, 10, 12 and 13) the annuli are scaled such
that for an $L_{*}$ galaxy, the width of each ring corresponds to a
physical scale of $\sim$ 20 kpc at  $z = 0.175$. 

\subsection{Error Estimate on the signal}

There are two principal sources of error in the computation of the
averaged value of the shear aside from
the observational errors (which are not taken into account in these
simulations) arising from the effects of seeing etc. (i) shot noise
(due to a finite number of sources and the intrinsic width of their 
ellipticity distribution) and (ii) in principle the unknown source
redshifts. Therefore, we require a minimum threshold number of
background objects to obtain a
significant level of detection. The unknown redshift
distribution of the sources also introduces noise and affects the
retrieval of the signal in a systematic way, for instance, the
obtained absolute value for the total mass estimate for cluster
galaxies is an under-estimate for a higher redshift population for a 
given measured value of the shear. 
The mean (or alternatively the median) and width of the redshift
distribution are the important
parameters that determine the errors incurred in the extraction
procedure. 

For the simulation however, we need to obtain an
error estimate on the signal given that we measure the averaged shear for a 
single realization. In order to do so, the simulation was set up with 
a constant mass-to-light ratio ($\Upsilon\,=\,$12) for the 50 cluster 
galaxies with 5000 background galaxies, and on solving the lens
equation the image frame was obtained. The averaging procedure as
outlined in Section 2.1 was then implemented to extract the output 
signal with 1000 independent sets of random scrambled positions for
the cluster galaxies (in addition to the one set of 50 positions that
was actually used to generate the image); the results are plotted as 
the lower solid curves in Figures 7 \& 8. This is a secure
estimate of the error arising for an individual realization, since
this error arises primarily from the dilution of the strength of the
measured shear due to uncorrelated sources and lensed images. We found that the 
mean error in $\left<{\tau_{u}}\right>$
in the first annulus is $0.040\pm0.0012$ and $0.0048\pm0.0047$ in
$\left<{\tau_{v}}\right>$. 

\subsection{Variation of the signal with mass-to-light ratio of
cluster galaxies}

The simulations were performed for mass-to-light ratios ($\Upsilon$)
ranging from 2 $\to$ 48 (see Figures 6, 7 \& 8). The velocity
dispersion of the fiducial galaxy model was adjusted to 
give the requisite value for $\Upsilon$ keeping the scaling relations 
intact. The detection is significant for mass-to-light ratios 
$\Upsilon\,\geq\,$4 given the configuration with 50 cluster galaxies
and 5000 background galaxies. The strength  of the signal varies with the 
input $\Upsilon$ of the cluster galaxies, and increases with 
increasing $\Upsilon$. As a test run, with $\Upsilon\,=\,0$, (i.e. no 
cluster galaxies) and only the large-scale component of the shear, we
do recover the expected behavior for $\left<{\tau_{u}}\right>$. The signal was
extracted for both background source redshift distributions MODEL Z1 \& 
MODEL Z2. While the amplitude of the signal is not very sensitive to 
the details of the redshift distribution of the background population 
and hence did not vary significantly, the error-bars are marginally 
larger for MODEL Z2.  This can be understood in terms of the
additional shot noise induced due to the change in
the relative number of objects `available' for lensing; in MODEL Z2 a 
fraction of the galaxies in the low-z tail of the redshift
distribution end up as foreground 
objects and are hence not lensed, thereby diluting the signal and
increasing the size of the error-bars marginally.
\begin{figure}
\psfig{figure=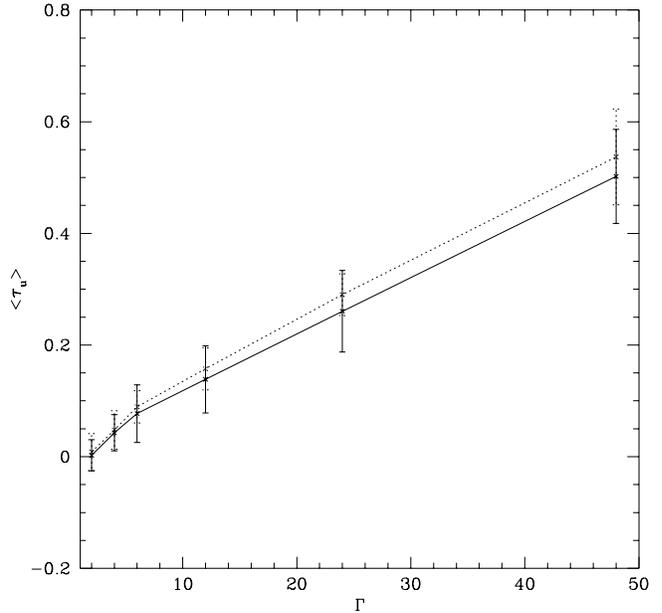,width=0.5\textwidth}
\caption{Variation of the mean value of the signal in the first
annulus with mass-to-light ratio $\Upsilon$: for $\Upsilon$= 2, 4, 6, 
12, 24, 48 of the
cluster galaxies plotted for MODEL Z1 (solid curve) and MODEL Z2
(dotted curve).}
\end{figure}
\begin{figure}
\psfig{figure=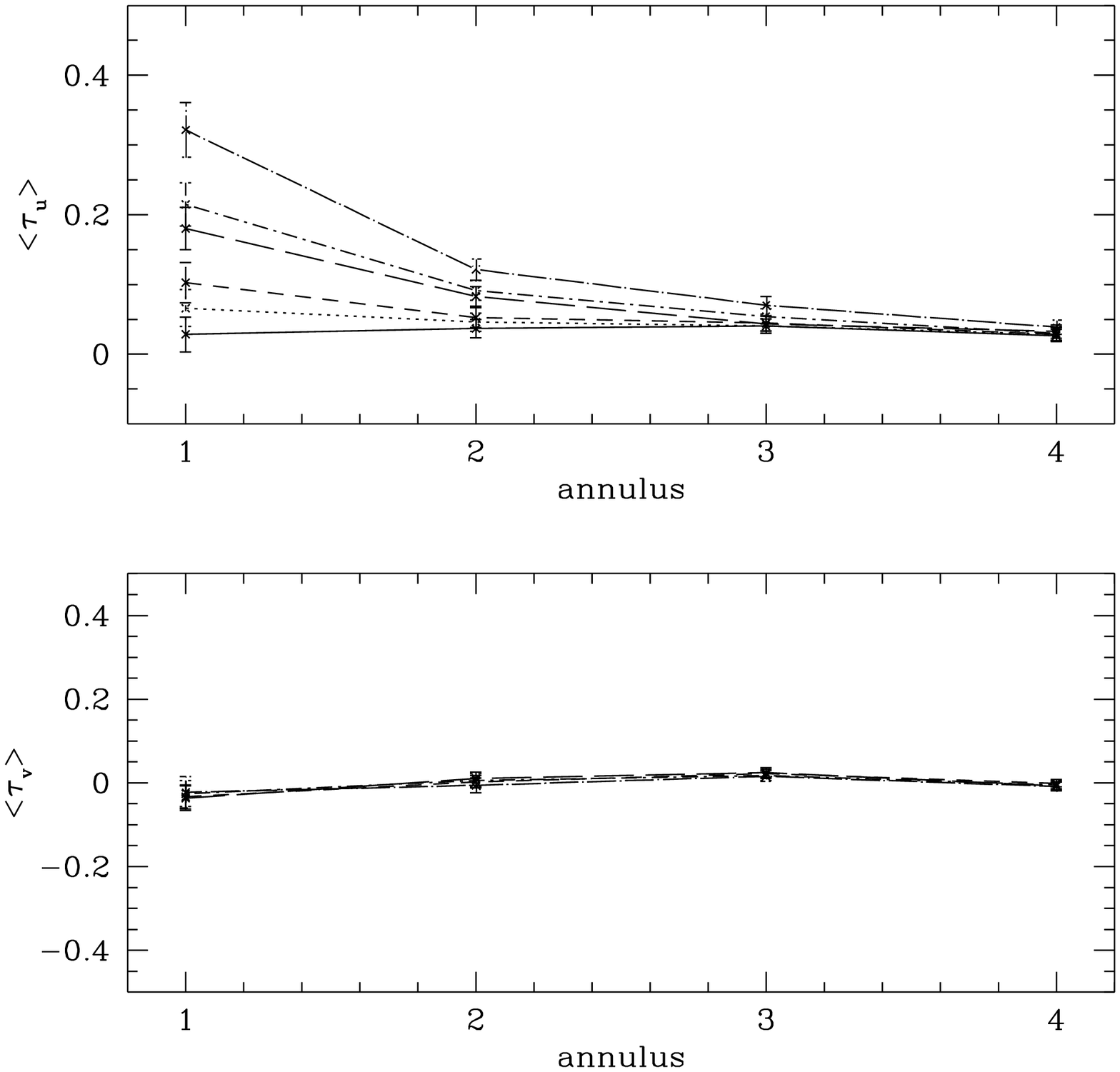,width=0.5\textwidth}
\caption{Recovering the  signal for MODEL Z1: for various 
values of the constant mass-to-light ratio $\Upsilon$ of the cluster galaxies 
ranging from 2 - 48, (i) lower solid curve - the error estimate  (ii) upper 
solid curve - $\Upsilon$ = 2, (iii) dotted curve - 
$\Upsilon$ = 4, (iv) dashed curve - $\Upsilon$ = 6, (v) long-dashed curve - 
$\Upsilon$ = 12, (vi) dot-short dashed curve - $\Upsilon$ = 24, 
(vii) dot-long dashed curve - $\Upsilon$ = 48. Note here that
$<{\tau_v}>$ is zero as expected by definition of the (u,v) coordinate
system.}
\end{figure}

\begin{figure}
\psfig{figure=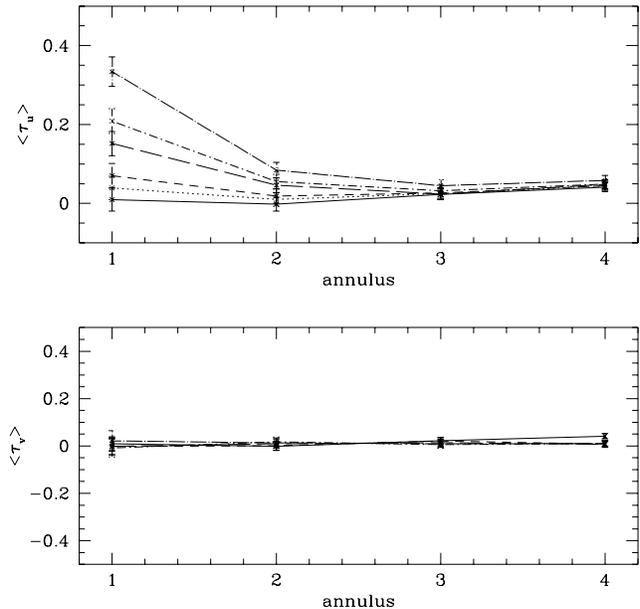,width=0.5\textwidth}
\caption{Recovering the signal for MODEL Z2: for various values of the
  constant mass-to-light ratio $\Upsilon$ of the cluster galaxies
  ranging from 2 - 48,
(i) lower solid curve - the error estimate  (ii) upper 
solid curve - $\Upsilon$ = 2, (iii) dotted curve - 
$\Upsilon$ = 4, (iv) dashed curve - $\Upsilon$ = 6, (v) long-dashed curve - 
$\Upsilon$ = 12, (vi) dot-short dashed curve - $\Upsilon$ = 24, 
(vii) dot-long dashed curve - $\Upsilon$ = 48.}
\end{figure}

\subsection{Variation with the number of background galaxies}

The efficiency of detection of the signal depends primarily on the
number of background galaxies averaged over in each annulus and
therefore on the number that are lensed by the individual
cluster galaxies. For a fixed value of $\Upsilon$, the total
number of background galaxies $N_{bg}$ was varied, assuming a redshift
distribution of the form of MODEL Z1. With increasing $N_{bg}$, 1000 $\to$
2500 $\to$ 5000, the detection is more secure and the error does vary 
roughly as $\sqrt{N_{bg}}$ as shown in Figure 9. In principle, the larger 
the number of background sources available for lensing, the more 
significant the detection with tighter error bars: however we find 
that a ratio of 50 cluster galaxies to 2500 background galaxies
provides a secure detection for $\Upsilon\,\geq\,4$, a larger number of background
source are required to detect the corresponding signal induced by lower
mass-to-light ratio halos. A secure detection in this case refers to
the fact that the difference in the mean values of the detected signal
in the two cases (with $N_{bg}$ = 5000 and $N_{bg}$ = 2500 background sources) is
comparable to the mean estimated error per realization computed in
Section 4.1. The number count distribution used to generate the
background sources corresponds to a background surface number density
of $\sim$ 90 galaxies per square arcmin which we find provides a
secure detection for $\Upsilon\,\geq\,4$. It is useful to point out
here that for the standard Bruzual \& Charlot (95) spectral evolution
of stellar population synthesis models with solar metallicity, a galaxy that is roughly
10 Gyr old (a reasonable age estimate for a galaxy in a $z\,\sim\,0.3$
cluster), formed in a single 1Gyr burst of star formation and
having evolved passively, one obtains a stellar mass-to-light ratio
in the R band of $\sim 8$ with a single power law Salpeter IMF with 
lower mass limit of $0.1\,M\odot$ and upper mass limit $125\,M\odot$.
With the same ingredients but a Scalo IMF one obtains a M/L ratio
about a factor 2 smaller ($\sim 4$) since there are a smaller proportion of
very low-mass stars. Therefore, an R-band M/L of 4 for a cluster
galaxy is consistent with the observed mass just in stars and does not
imply the presence of any dark mass in the system. Therefore, if dark
halos were indeed present around the bright cluster members, the
corresponding inferred mass-to-light ratios would be greater than 4,
and with 5000 background galaxies, we would be sensitive to the signal
as shown in the plots of Figures 6, 7 \& 8.   
\begin{figure}
\psfig{figure=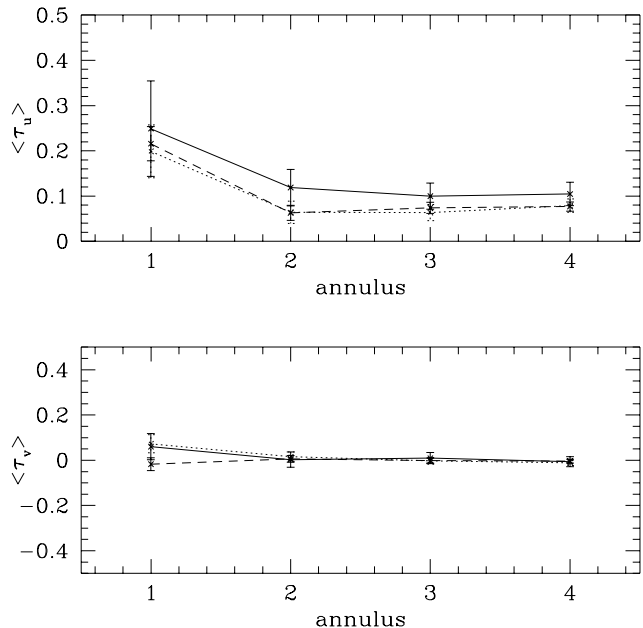,width=0.5\textwidth}
\caption{Variation of the signal with the number of background
galaxies for MODEL Z1:
for a given mass-to-light ratio $\Upsilon$= 12 of the cluster
galaxies. We find that the error bars and hence the noise decrease 
as expected with increasing $N_{bg}$ (i) solid curve: $N_{bg}$ = 1000,
 (ii) dashed curve: $N_{bg}$ = 2500, (iii) dotted curve: $N_{bg}$ = 5000.}
\end{figure}

\subsection{Variation with cluster redshift}

The lensing signal depends on the distance ratio $D_{ls}/D_{os}$,
the angular extent of the lensing objects, the number
density of faint objects and their redshift distribution.
We performed several runs with the cluster
(the lens) placed at different redshifts, ranging from $z\,=\,$0.01 to 0.5.
We scaled all the distances with the appropriate factors corresponding 
to each redshift for both MODELS Z1 \& Z2. For MODEL Z1 (Figure 10 and
dotted curve in Figure 12), we
find that the signal (by which we refer to the 
value of $\left<{\tau_{u}}\right>$ in the innermost annulus)
saturates at low redshifts; for $0.01\,<\,z_{\rm lens}\,<\,0.07$ the 
measurements are consistent with no detection but the strength 
increases as $z_{\rm lens}$ is placed further away and it remains 
significant for upto $z_{\rm lens}\,=\,$0.4, subsequent to which it 
falls sharply once again at 0.5. On the other hand, we find that for 
MODEL Z2 (Figure 11, and solid curve in Figure 12),
there is a well-defined peak 
and hence an optimal lens redshift range for extracting the signal. Thus,
in general, cluster-lenses lying between redshifts 0.1 and 0.3 are
the most suitable ones for constraining the mean $\Upsilon$ of
cluster galaxies via this direct averaging procedure. These trends
with redshift can be understood easily,  
the shear produced is proportional to the surface mass
density and scales as ($D_{\rm ls}/D_{\rm os}$) - the saturation at
high-redshift is due to the combination of two diluting effects (i)
the decrease in $D_{\rm ls}$  as the lens is placed at successively
higher redshifts (ii) the effect of
additional noise induced due to a reduction in the number of
background objects for MODEL Z2. 
The drop-off at low z (for both models) is
primarily due to behavior of the angular scale factors at low-redshifts.
Additionally, the shape of these curves is independent of the total
mass of the cluster (the total mass being dominated by the smooth
component), therefore even for a subcritical cluster we obtain the same
variation with redshift.
\begin{figure}
\psfig{figure=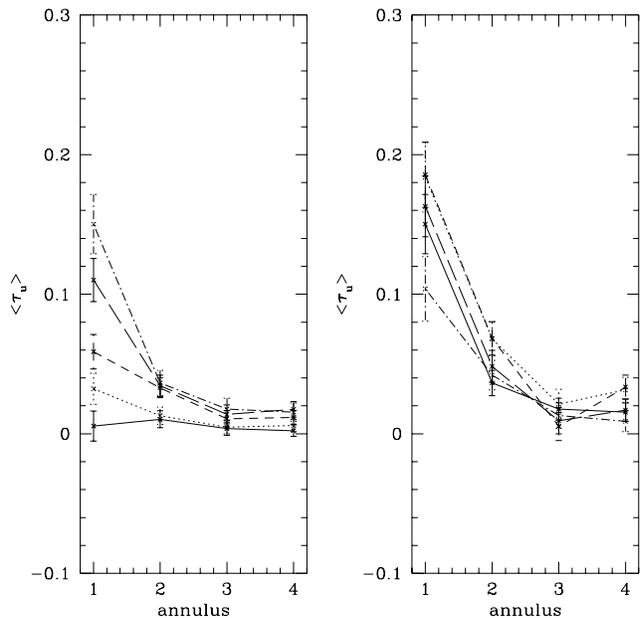,width=0.5\textwidth}
\caption{Variation of the signal with cluster redshift for MODEL Z1:
for a given mass-to-light ratio $\Upsilon$= 12 of the cluster galaxies 
placing the lens at different redshifts, right panel:(i) solid curve: z =
0.1, (ii) dotted curve: z = 0.2, (iii) dashed curve: z
= 0.3, (iv) long dashed curve: z = 0.4, (v) dot dashed 
curve: z = 0.5 and the left panel:(i) solid curve: z = 0.01, (ii) dotted
curve: z = 0.02, (iii) dashed curve: z = 0.05, (iv) long dashed 
curve: z = 0.07, (v) dot dashed curve: z = 0.10}
\end{figure}
\begin{figure}
\psfig{figure=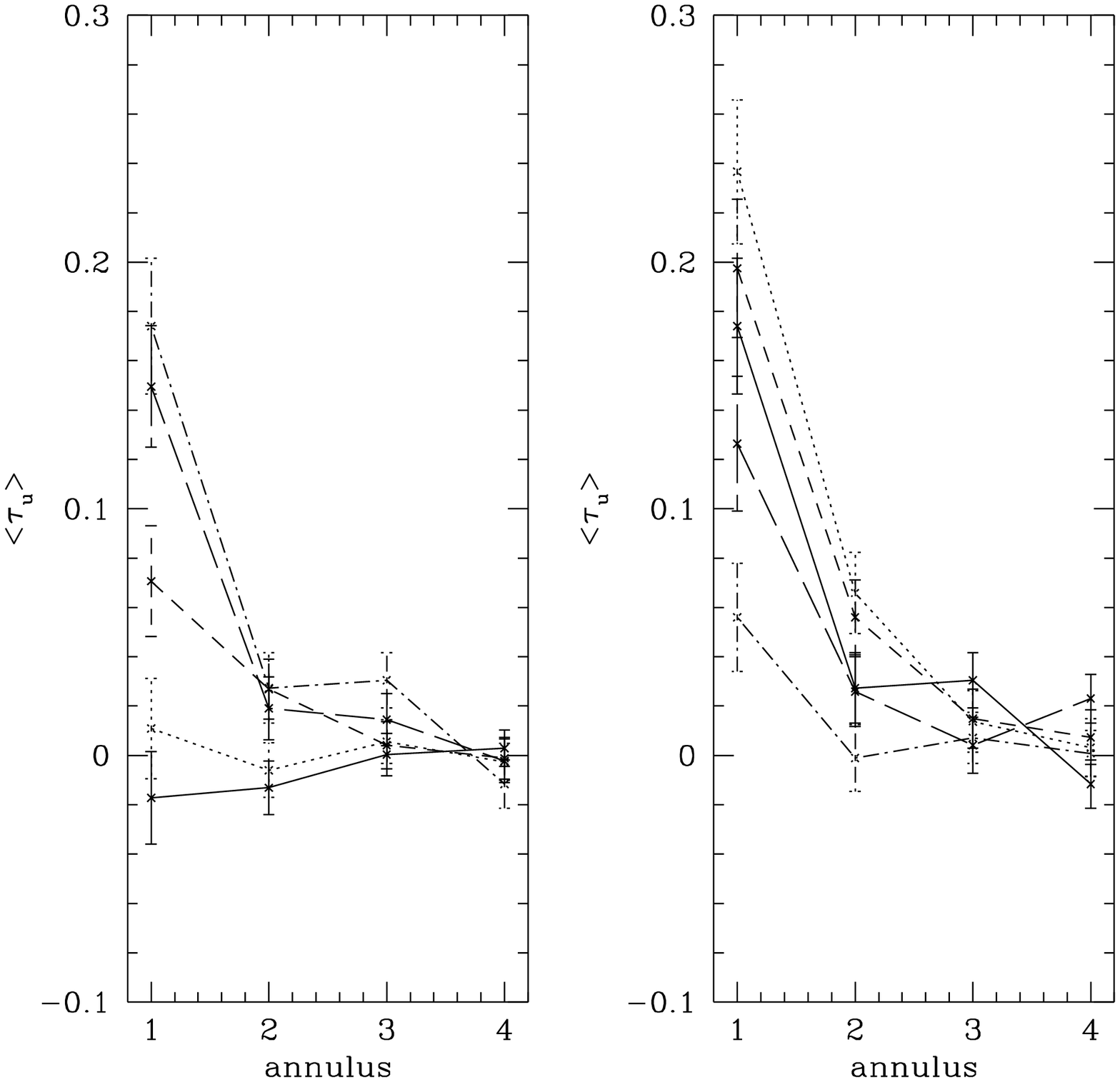,width=0.5\textwidth}
\caption{Variation of the signal with cluster redshift for MODEL Z2:
for a given constant mass-to-light ratio $\Upsilon$= 12 of the cluster
galaxies placing the lens at different redshifts, right panel (i) solid curve: 
z = 0.1, (ii) dotted curve: z = 0.2, (iii) dashed curve: z
= 0.3, (iv) long dashed curve: z = 0.4, (v) dot dashed 
curve: z = 0.5; left panel (i) solid curve: z = 0.01, (ii) dotted
curve: z = 0.02, (iii) dashed curve: z = 0.05, (iv) long dashed 
curve: z = 0.07, (v) dot dashed curve: z = 0.10 }
\end{figure}
\begin{figure}
\psfig{figure=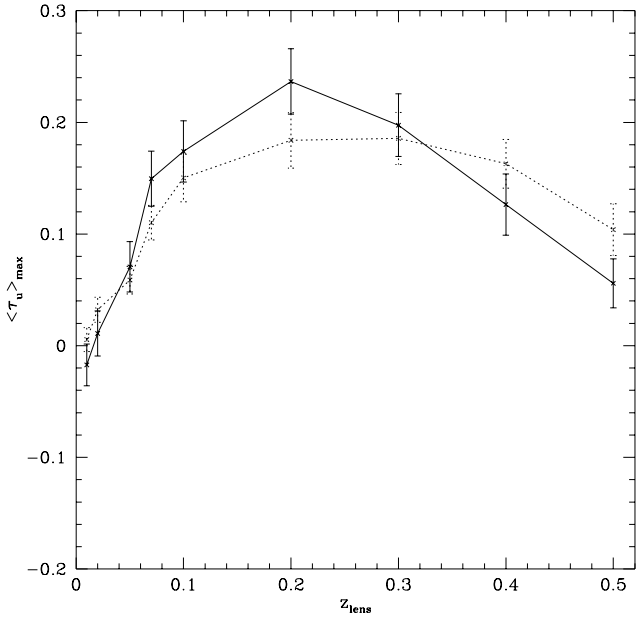,width=0.5\textwidth}
\caption{Variation of the maximum value of the signal with redshift:
for a given constant mass-to-light ratio $\Upsilon$= 12 of the cluster
galaxies placing the lens at different redshifts for the 2 background
redshift distributions for the sources (i) dotted curve: MODEL Z1,
(ii) solid curve: MODEL Z2}
\end{figure}

\subsection{Dependence on assumed scaling laws}

In section 3.1, we outlined the simple scaling relations that were used 
to model the cluster galaxies. The choice of the exponent $\alpha$ in
equation (26) allows the modelling of the trends for different galaxy
populations: $\alpha\,=\,$0.5 corresponding to a constant $\Upsilon$ and
$\alpha\,=\,$0.8 corresponding to $\Upsilon$ being a weak function of 
the luminosity.
\begin{figure}
\psfig{figure=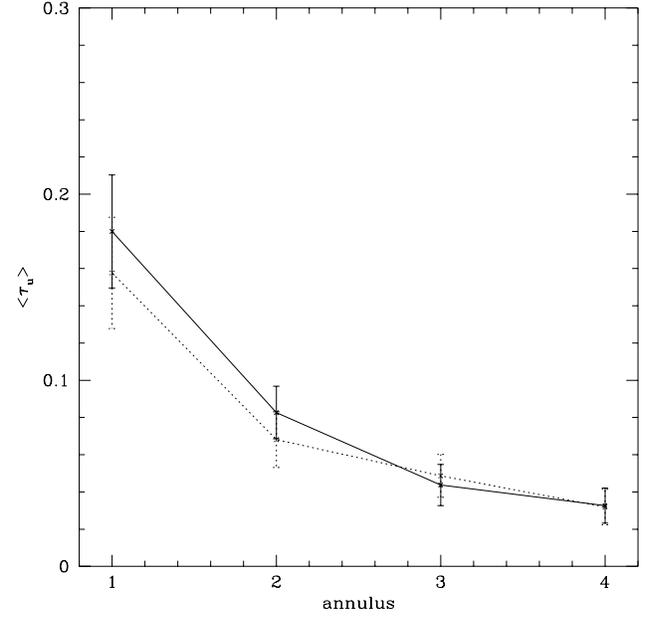,width=0.5\textwidth}
\caption{Examining the scaling relations - the 2 choices of 
$\alpha$ the exponent of the scaling 
relation for the truncation radius for $\Upsilon$= 12. We plot the 
recovered signal (i) solid curve: $\alpha$ = 0.8,  
(ii) dotted curve: $\alpha$ = 0.5.}
\end{figure}
Simulating both these cases above, we find that the mean value
of the signal does depend on the assumed exponent for the scaling law 
and therefore the mass enclosed (Figure 13). 
We find that while the signal is stronger 
for $\alpha = 0.8$, since that corresponds to the mass-to-light 
ratio ${\Upsilon}\,\sim\,({L \over {L_*}})^{0.3}$ - on average that is
what we expect compared to the constant mass-to-light ratio
case; it is not possible however, to distinguish between a
correspondingly higher value of the constant M/L and a higher value of 
$\alpha$. Therefore, the direct averaging procedure cannot
discriminate between the assumed detail 
fall-off for the mass distribution. 

\subsection{Examining the assumption of analysis in the weak regime}

While our mathematical formulation outlined in section 2 is strictly
valid only for $\kappa\,\ll\,1$, we examine how crucial this assumption
is to the implementation of the technique. For the output images from 
the simulations, the magnification $\kappa$ is known at all points.
Prior to the averaging, we excised the high $\kappa$ regions
successively, by removing only the lenses in those regions.
The results are plotted in Figure 14 for input 
$\Upsilon$ = 12, with the sources distributed as
specified by MODEL Z1. While the mean peak value of the signal does
not fluctuate much; on removing the high-$\kappa$ regions, we find
that the cluster subtraction does get progressively more efficient, as
evidenced by the sharp fall-off to zero of the signal in the second
annulus outward. Therefore, while the detectability and magnitude of
the signal is robust even in the `strong regime', the contribution
from the smooth cluster component, which for our purposes is a
contaminant, can be `removed' optimally only in the low $\kappa$ regions.
\begin{figure}
\psfig{figure=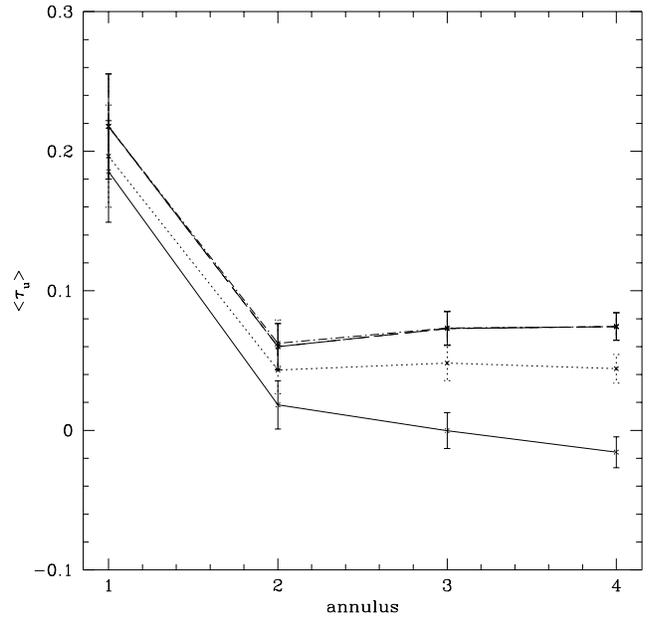,width=0.5\textwidth}
\caption{The effect of excising the high $\kappa$ regions in the image:
(for $\Upsilon=12$ of the cluster galaxies)
(i) solid curve: $\kappa \leq$ 0.1, (ii) dotted curve:$\kappa \leq$
0.2, (iii) dashed curve:$\kappa \leq$ 0.3, (iv) long dashed curve: 
$\kappa \leq$ 0.4, (v) dot dashed curve:$\kappa \leq$ 0.5.} 
\end{figure}

\section{Maximum-Likelihood Analysis}

\subsection{Limitations of the direct averaging method}

The simulations have enabled us to delineate the role of relevant 
parameters and comprehend the trends with cluster redshift,
the redshift distribution of the sources and the mass-to-light ratio
of the cluster galaxies. While the direct method to estimate the 
suffers from the following  limitations, specially in the cluster core,
(i) being in the `strong' lensing regime, the `cluster 
subtraction' is not very efficient and (ii) the probability of a 
background galaxy being sheared due to the cumulative effect of two 
or more cluster galaxies is enhanced; the core being a region with 
a high number density of cluster galaxies; it does, however, provide 
a robust estimate of the mass-to-light ratio modulo the assumed 
model parameters. 

We now explore applying a maximum-likelihood method 
to obtain significance bounds on fiducial parameters that characterize 
a `typical' galaxy halo in the cluster. \citeN{schneiderrix96} developed a
maximum-likelihood prescription for galaxy-galaxy lensing in the
field; here we develop one to study lensing by galaxy halos embedded in the cluster.  
Schematically, we demonstrate the differences in the ellipticity
distribution that we are attempting to discern in Figure 15.  Here we
have plotted the intrinsic ellipticity distribution of the unlensed
sources, sources lensed only by a cluster scale component 
and sources sheared by both a cluster scale component and 50
cluster galaxies; from which it is obvious that the effect that we
intend to measure in terms of parameters that characterize the cluster 
galaxies is indeed small, hence recovery of the fiducial parameters in
this case is considerably harder than in the case of purely galaxy-galaxy lensing.
\begin{figure}
\psfig{figure=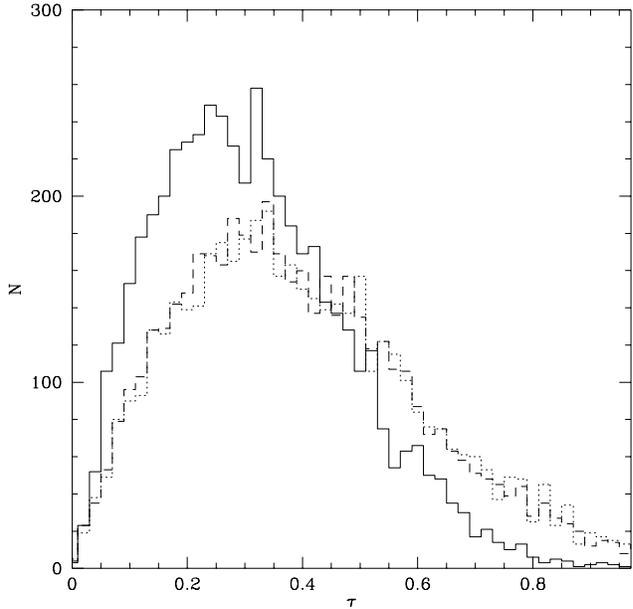,width=0.5\textwidth}
\caption{The ellipticity distribution $p_{\tau_S}$: (i) solid curve - intrinsic
input ellipticities of the sources, (ii) dotted curve - the
ellipticity distribution on being lensed by 50 galaxy-scale mass
components and one larger-scale smooth component and (iii) dashed
curve - the ellipticity distribution induced by lensing 
only by the larger-scale smooth cluster component.}
\end{figure}

\subsection{Application of the maximum-likelihood method}

The basic idea is to maximize a likelihood function of the estimated
probability distribution of the source ellipticities for a set of
model parameters, given the functional form of the intrinsic
ellipticity distribution measured for faint galaxies.
We briefly outline the exact procedure below. From the simulated image
frames we extract the observed ellipticity $\tau_{\rm obs}$. 
For each `faint' galaxy $j$, the source ellipticity can then be 
estimated  {\sl in the weak regime} by just subtracting the lensing 
distortion induced by the smooth cluster and galaxy halos given the
parameters that characterize both these mass distributions, in other
words,
\bea
\tau_{S_j} \,=\,\tau_{\rm obs_j}\,-{\Sigma_i^{N_c}}\,{\gamma_{p_i}}\,-\,
 \gamma_{c},
\eea
where $\Sigma_{i}^{N_{c}}\,{\gamma_{p_i}}$ is the sum of the shear
contribution at a given position $j$ from $N_{c}$ perturbers, and the 
term $\gamma_{c}$ is the shear induced by the smooth cluster component. 
In the strong regime, similarly, one can compute the source ellipticity
using the inverse of equation (7). The lensing distortion depends on 
the parameters of the smooth cluster potential, the perturbers and on 
the redshift of the observed arclet (lensed image), which is in general
unknown. Therefore, in order to invert equation (7), for each lensed galaxy
we need to assign a redshift, from a distribution of the form
in equation (33) given the observed magnitude $m_{j}$ and take the mean of
many such realizations. In principle, one needs to also
correct the observed magnitude for amplification to obtain the true
magnitude prior to drawing a redshift from $N(z,m)$, but this
correction in turn depends on the redshift as well. An alternative
procedure is then to correct for the amplification using the median $z$
corresponding to the observed magnitude from the same distribution. 
This entire inversion procedure is performed
within the lens tool utilities, which accurately takes into account
the non-linearities arising in the strong regime. As an input for this
calculation, we parameterize both the large-scale component and
perturbing galaxies as described in Section 2.2 and Section 3.1
respectively. Additionally, we assume that a well-determined `strong 
lensing' model for the cluster-scale halo is known.
For our analysis, we also assume that the functional form of 
$p(\tau_{S})$ from the field is known, and is specified by equation (34);
the likelihood for a guessed model can then be expressed as,
\begin{eqnarray}
 {\cal L}({{\sigma_{0*}}},{r_{t*}}, ...) = 
\Pi_j^{N_{gal}} p(\tau_{S_j}).
\end{eqnarray}
However, note that we ought to compute ${\cal L}$ for different
realizations of the drawn redshift for individual images (say about
10-20) and then compute the mean of the different realizations of
${z_{j}}$; but it is easily shown to be equivalent to constructing
the ${\cal L}$ for a single realization where the redshift ${z_{j}}$
of the arclet drawn is the median redshift corresponding to the
observed source magnitude. For the case when we perform a Monte-Carlo sum over $N_{\rm
MC}$ realizations of $z_j$, the likelihood is:
\begin{eqnarray}
{\cal L}({{\sigma_{0*}}},{r_{t*}}, ...) = 
\Pi_j^{N_{gal}} {\Pi_k^{N_{\rm MC}}}p(\tau_{S_j^k}),
\end{eqnarray}
where $p_{\tau}(\tau_{S_j^k})$ is the probability of the source
ellipticity distribution at the position $j$ for $k$ drawings
for the redshift of the arclet of known magnitude $m_j$. The mean
value for $N_{\rm MC}$ realizations gives:
\begin{eqnarray}
\left<{p(\tau_{S_j})}\right>\,=\,{1 \over {N_{\rm
MC}}}\,{\Sigma_{k=1}^{N_{\rm MC}}}\,{p(\tau_{S_j^k})}
\end{eqnarray}
which written out in integral form is equivalent to 
\begin{eqnarray}
\left<{p(\tau_{S_j})}\right>\,&=&\,{{\int {{p(\tau_{S_j}(z))}{N(z,{m_j})\,dz}}} \over
{\int {N(z,{m_j})\,dz}}}\,\\ \nonumber &=&\,{p(\tau_{S_j}({z_{\rm avg}}))}\,\sim\,{p(\tau_{S_j}({z_{\rm median}}))}
\end{eqnarray} 
${z_{\rm avg}}$ being the average redshift corresponding to the magnitude
$m_j$.
Therefore the corresponding likelihood ${\cal L}$ is then simply,
\begin{eqnarray}
{\cal L}\,=\,{\Pi_j}{\left<{p({\tau_{S_j}})}\right>}
\end{eqnarray}
as before and the log-likelihood $l\,=\,\ln {\cal
L}\,=\,{\Sigma}{\left<{p(\tau_{S_j})}\right>}$.
The best fitting model parameters are then obtained by maximizing this
log-likelihood function $l$ with respect to the parameters
${\sigma_{0*}}$ and ${r_{t*}}$, the characteristic central velocity
dispersion and truncation radius respectively. The results of the
maximization are presented in Figures 16 - 18. For all reasonable choices of
input parameters we find that the log-likelihood function has a
well-defined and broad maximum (interior to the innermost contour  on
the plots). The contour levels are calibrated such that ${l_{\rm
max}}\,-\,l\,=\,1, 2, 3$ can be directly related to confidence
levels of 63\%, 86\%, 95\% respectively (we plot only the first 10 
contours for each of the cases in Figures 16 - 18) and the value marked by the
dotted lines denotes the input values. In Figure 16, we plot the 
likelihood contour for the MDS ellipticity distribution (equation
(34)) - the left panel for an assumed scaling law with $\alpha$ =
0.5 and a constant mass-to-light ratio $\Upsilon$ = 12. On the right
panel, the corresponding contours for $\alpha$ = 0.8 are plotted. For the MDS
ellipticity distribution, we find that the velocity dispersion ${\sigma_{0*}}$
can be more stringently constrained than the halo size, and
the contours are elongated along the constant mass-to-light ratio curves
and yield an output $\Upsilon$ very nearly equal to the input value. For narrower
ellipticity distributions both the parameters can be constrained
better and the inferred $\Upsilon$ is every nearly equal to the input
value. We find that there is very little perceptible difference in
the retrieval of parameters for the two cases with the different scaling laws. 
For a sub-critical cluster (see bottom left panel in Figure 18), we find that the
parameters are recovered just as reliably, which is not surprising and in
some sense illustrates the robustness of the maximum-likelihood method.
Thus, the physical quantity of interest that can be estimated best
from the analysis above is the mass $M_*$ of a fiducial $L_*$ galaxy.

\begin{figure*}
\psfig{figure=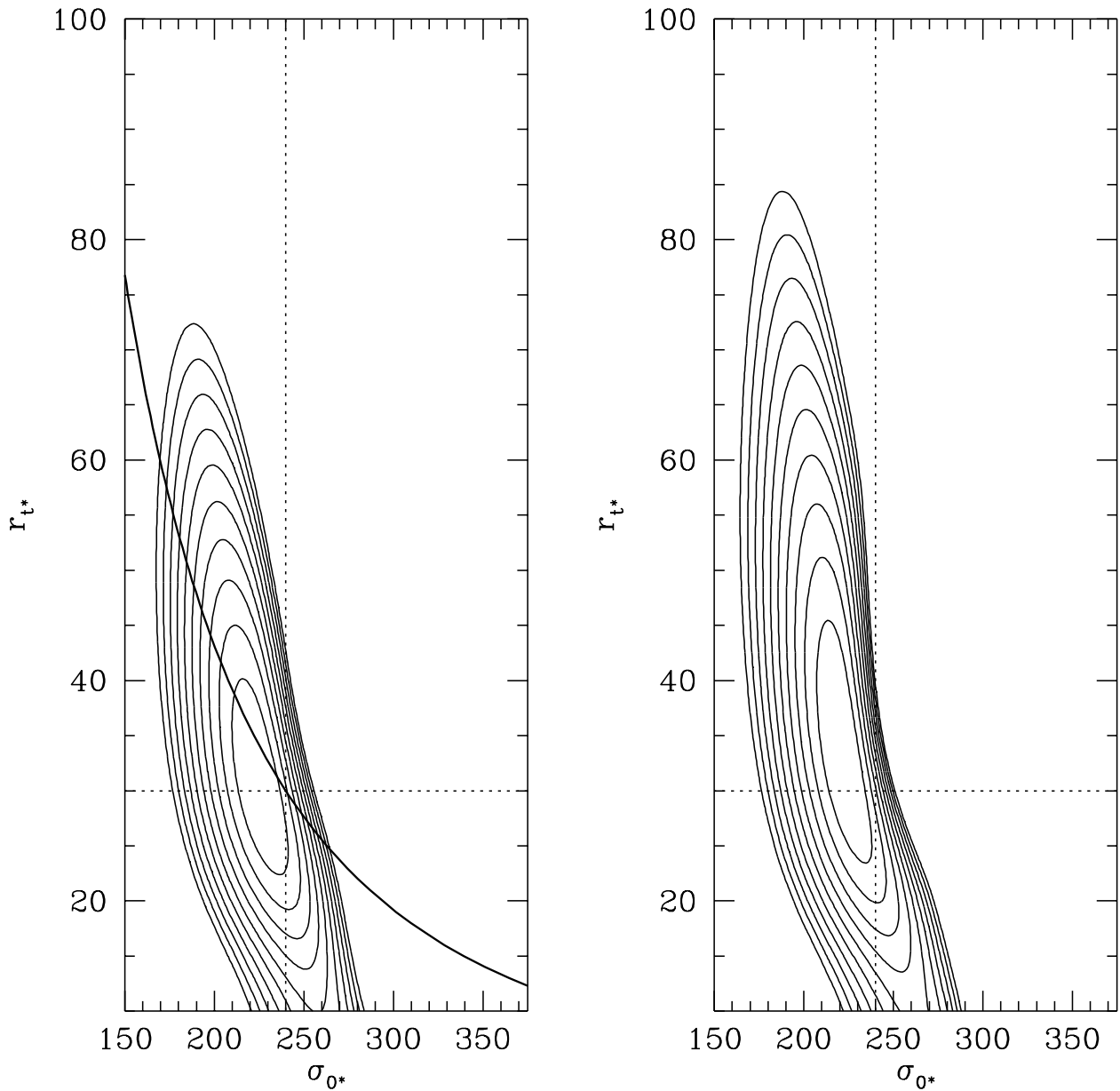,width=1.0\textwidth}
\caption{Log-likelihood contours for the retrieval of the fiducial
parameters $\sigma_{0*}$ and $r_{t*}$ - the input values are indicated
by the intersection of the dashed lines. In the left panel: For
the MDS ellipticity distribution, with assumed scaling $\alpha$ = 0.5, right
panel: the same with $\alpha$ = 0.8}
\end{figure*}

\begin{figure*}
\psfig{figure=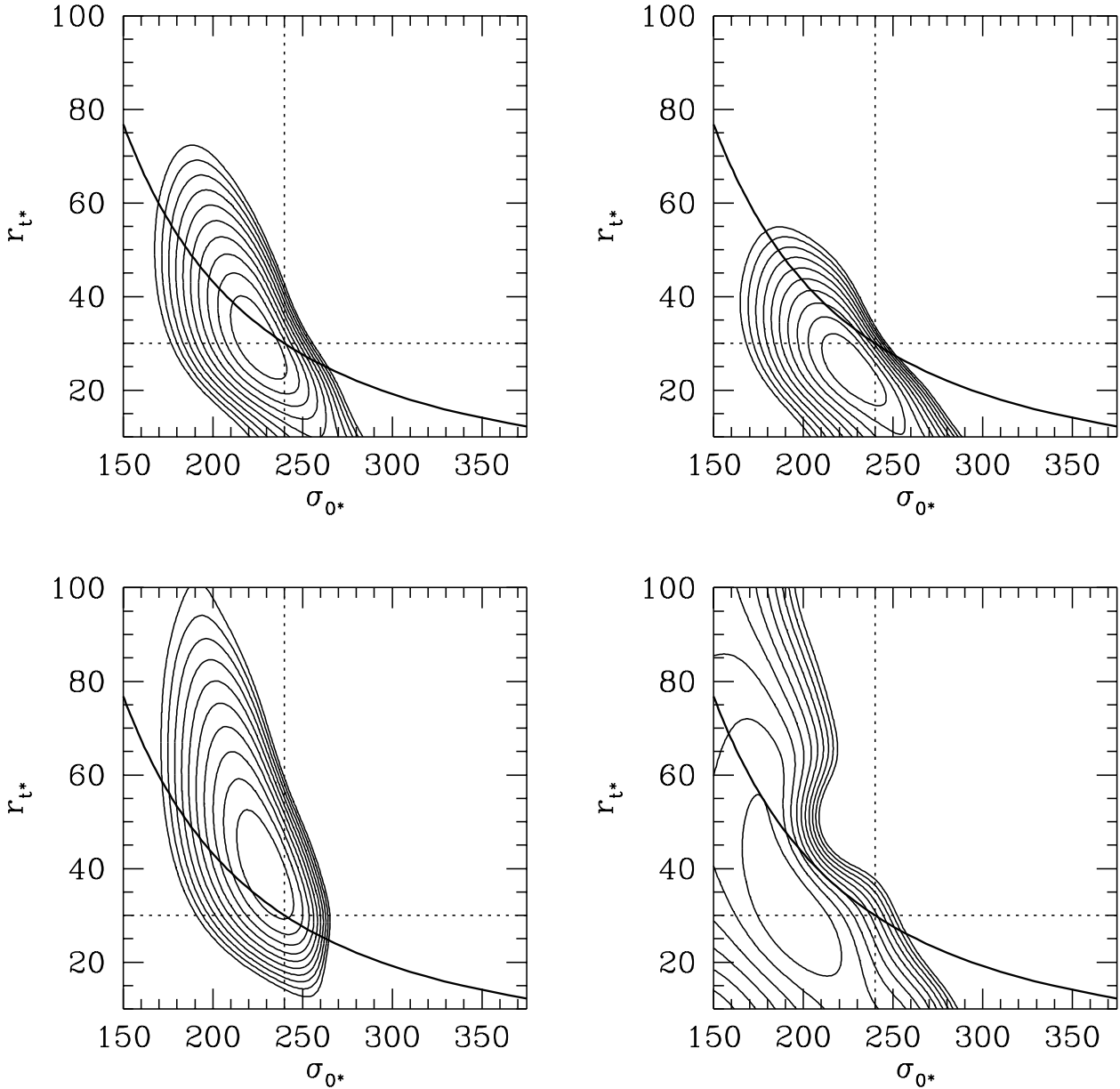,width=1.0\textwidth}
\caption{Sensitivity of log-likelihood contours to the strong lensing
input parameters: examining the tolerance of the
significance bounds obtained on $\sigma_{0*}$ and $r_{t*}$ with regard to the 
accuracy with which the cluster velocity dispersion needs to be
known. All plots are for input $\Upsilon$ = 12, $\alpha$ = 0.5 and the
MDS source ellipticity distribution. Top left panel: given that
the exact value of the velocity dispersion is known (value in this
case is 1090 km$s^-1$), top right panel: the velocity dispersion 
known to within 2\%, bottom left panel: attempt to
retrieve the incorrect scaling law - input $\alpha$ = 0.5,
log-likelihood maximized for $\alpha$ = 0.8, bottom right panel:
retrieval with fewer background galaxies}
\end{figure*}

\begin{figure*}
\psfig{figure=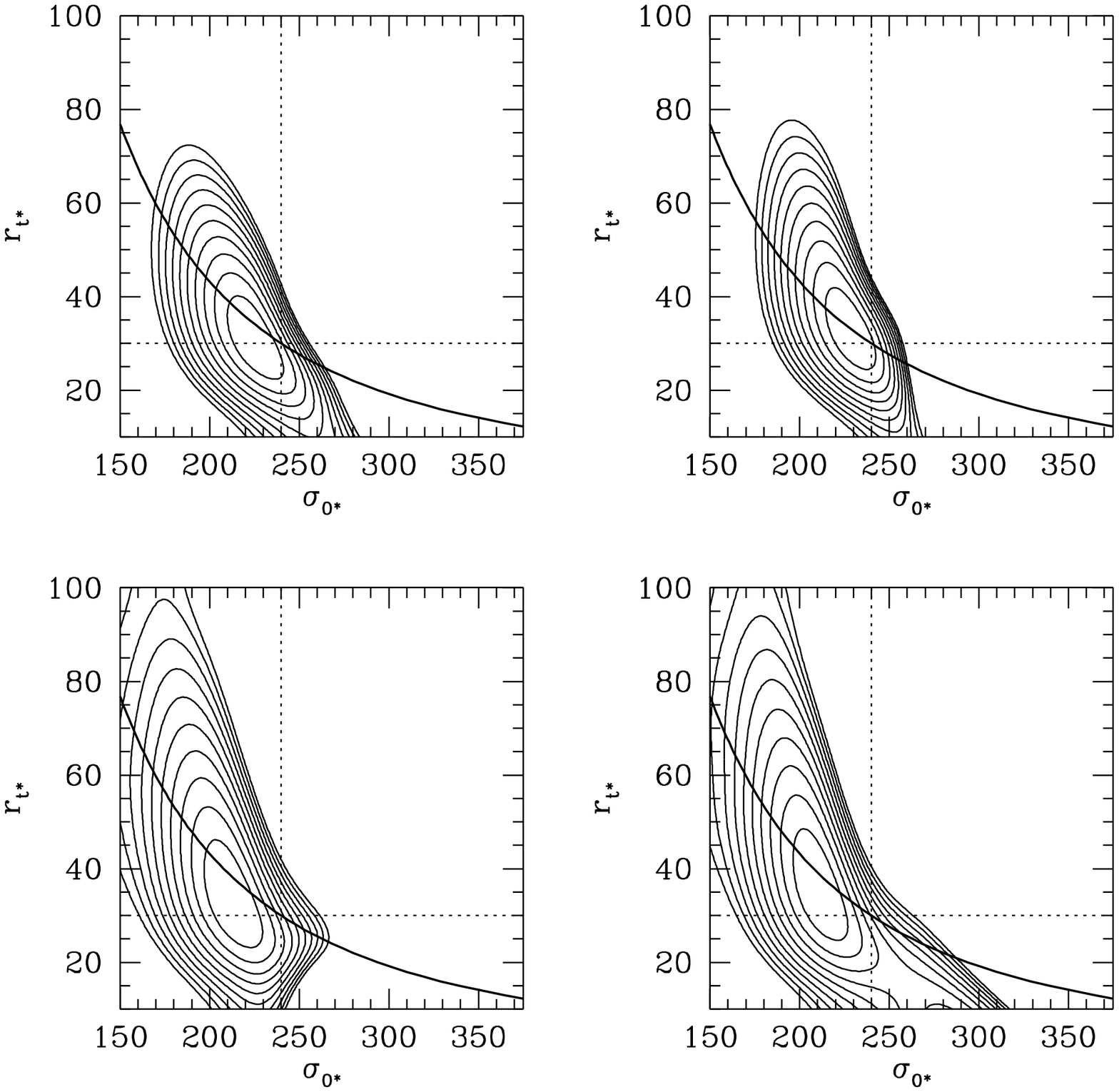,width=1.0\textwidth}
\caption{Sensitivity of the log-likelihood contours to input
parameters: examining the tolerance of the
significance bounds obtained on $\sigma_{0*}$ and $r_{t*}$ given the  
accuracy to which the cluster center needs to be
known. All plots are for input $\Upsilon$ = 12, $\alpha$ = 0.5 and the
MDS source ellipticity distribution. Top left panel: knowing the cluster
center exactly for the critical cluster, top right panel: knowing
the center to within 5 arcsec, bottom left panel: knowing the
center exactly for the sub-critical cluster and the bottom right
panel: for the subcritical cluster, center known to within 5 arcsec.}
\end{figure*}

\subsection{Estimating the required number of background galaxies}

The largest source of noise in our analysis arises due to the finite number of
objects in the frame. To estimate the required signal-to-noise that would permit
obtaining reliable constraints on both $\sigma_{0*}$ and $r_{t*}$, we reduced the
number of background sources to 2500 keeping the number of lenses at
50 as before. We do not converge to a maximum in the log-likelihood, 
 and consequentially, no confidence limits can be obtained on
the parameters. Therefore, to apply this technique to the data we require the
ratio $r$ of the number of cluster galaxies to the number of background
galaxies to be roughly $r\,<\,0.2$, which can be achieved only by
stacking the data from many clusters. Also, as found from the direct
averaging procedure, we require $\sim 5000$ lensed images in order to
securely detect $<\Upsilon>\,\geq\,4$. Although typical HST cluster data
fields are of the order of [3 arcmin X 3 arcmin]  have $\sim 700$
background galaxies (with a 10 orbit exposure) of these the shape parameters 
can be reliably measured only for about 200 galaxies, therefore on
stacking the data from 20 (10-orbit) HST cluster fields, we shall be
able to constrain statistically the mean mass-to-light ratios as well 
as the 2 fiducial parameters.  

\subsection{Uncertainties in the smooth cluster component}

In all of the above, we have assumed that the parameters that
characterize the smooth cluster-scale component are very accurately known
which is unlikely to be the case for the real data. We investigate the
error incurred in retrieving the correct input parameters from not
knowing this central strong lensing model well enough. So we can now
place limits on the order of magnitude of errors that can be tolerated
due to the lack of knowledge of the exact position of the cluster
center and the velocity dispersion of the main clump. In Figure 18,
we see that an uncertainty of the order of 20 arcsec in
the position of the center yields unacceptably incorrect values for $\sigma_{0*}$
and $r_{t*}$. Conversely, if the center is off by only 5 arcsec or so, for
both the critical cluster and the sub-critical one, the results remain
unaffected and we obtain as good a retrieval of the input $r_{t}^{*}$
as when the position of
the center is known exactly. Similarly, in Figure 17, we demonstrate
that an error of $\sim\,$5\% in the velocity dispersion is enough
to make the max-likelihood analysis inconclusive, but an error of $\sim$ 2-3\%
at most would still enable getting sensible bounds on both parameters. 

\section{\bf Conclusions and Prospects}

We conclude this section and assert that both the maximum-likelihood
method and the direct averaging method developed in this paper can be
feasibly applied to the real data on stacking a minimum of 20 WFPC2
deep cluster fields. These methods are well-suited to being used
simultaneously as they are
somewhat complementary, both yield the statistical mass-to-light ratio
reliably and while the averaging does not require either the knowledge
of the center or any details of the strong lensing model, it also
cannot provide the decoupling of the 2 fiducial parameters and hence no
independent constraints on the velocity dispersion and the
halo size can be obtained, meanwhile the maximum likelihood approach 
permits estimation of the fiducial $\sigma_{0*}$ and $r_{t*}$
($\sigma_{0*}$ more reliably than $r_{t*}$), it necessarily requires 
knowledge of the cluster center and the central velocity dispersion
rather accurately. In offset fields, however, where the gradient of
the smooth cluster potential is constant over the smaller scales that
we are probing, we expect both methods to perform rather well.

In this paper, we have not investigated the likely sources of error in the
real data, which we do in detail in a subsequent paper \cite{natarajan96a}, 
our simulations have enabled the study of the feasibility of 
application to HST cluster data in as far as a statistical estimate of
the required number of background galaxies required for a significant detection
given the limitations in the accuracy to which the input parameters (like
the strong lensing mass model and hence the magnification) are presently known.
Our analysis points to the fact that the extraction of the signal
would therefore be feasible if approximately 20 -- 25 clusters are
stacked, and the enterprise is specially suited to using the new ACS (advanced
camera for survey) due to be installed on HST in 1999. Additionally,
since there exists a well-defined optimum lens redshift for signal
detection (0.1$\,<\,{z_{\rm lens}}\,<\,$0.3), it might be useful to
target clusters in this redshift range in future surveys in order to
apply the techniques developed here. In our proposed analysis with the
currently available HST data, we intend to incorporate parameters
characterizing the smooth cluster (main clump) alongwith those of the 
perturbing galaxies into the maximum likelihood machinery.  

In summary, we have presented a new approach to infer the possible
existence of dark halos around individual bright galaxies in clusters
by extracting their local lensing signal. The composite lensing effect
of a cluster is modeled in numerical simulations via a large-scale
smooth mass component with additional galaxy-scale masses as
perturbers. The correct choice of coordinate frame i.e. the local frame
of each perturber, enables efficient subtraction of the shear induced
by the larger scale component, yielding the averaged shear field
induced by the smaller-scale mass component. Cluster galaxy halos
 are modeled using simple scaling relations and the background
high redshift population is modeled in consonance with observations
from redshift surveys. For several configurations of the sources and
lens, the lensing equation was solved to obtain the resultant images.
Not surprisingly, we find that the strength of the signal varies most
strongly with the mass-to-light ratio of the cluster galaxies and is
only marginally sensitive to the assumed details of the precise
fall-off of the mass profile. We also find that there is an optimum
lens redshift range for detection of the signal. Although the entire
procedure works in the `strong lensing' regime as well, it is less
noisy in the `weak regime'. The proposed maximum-likelihood method 
independently constrains the halo size and mass of a fiducial cluster
galaxy and we find that the velocity dispersion and hence the mass of a 
fiducial galaxy can be more reliably constrained than the characteristic halo size.
Examining the feasibility of application to real data, we find that
stacking $\sim$ 20 clusters allows a first attempt at extraction 
(\citeNP{natarajan96a}). The prospects for the application of this
technique are potentially promising,
specially with sufficient and high-quality data (either HST images or
ground-based observations under excellent seeing conditions of wider fields);
the mass-to-light ratios of the different morphological/color types in
clusters for instance can be probed. More importantly, comparing with similar
estimates in fields offset from the cluster center would allow us to
make the essential connections in order to understand the dynamical
evolution of galaxies in clusters and the re-distribution of dark
matter within smaller scales within clusters. Application of this
approach affords the probing of the structure of cluster
galaxies as well as the efficiency of violent  dynamical processes
like tidal stripping, mergers and  interactions which modify them and
constitute the processes by which clusters assemble. 

\section*{Acknowledgments}

PN thanks Martin Rees for his support and encouragement during the
course of this work. We acknowledge useful discussions with Alfonso
Aragon-Salamanca, Philip Armitage, Richard Ellis, Bernard Fort, Jens
Hjorth, Yannick Mellier, Ian Smail and Simon White. PN acknowledges 
funding from the Isaac Newton Studentship and Trinity College, JPK 
acknowledges support from an EC-HCM fellowship and from the CNRS.

% Now comes the reference list.  Since we typed out the citations ourselves,
% the reference list is enclosed in a "references" environment.  Each
% new reference begins with a \reference command which sets up the proper
% indentation.  Typography that may be required in the reference list by
% the editorial staff must be included by the author.
%
% Observe the "standard" order for bibliographic material: author name(s),
% publication year, journal name, volume, and page number for articles.
% Some journal names are available as macros; see the WGAS markup
% instructions for a listing of which ones have been "macro-ized".
% Note the use of curly braces to delimit the font changes: it is essential
% that this be done to limit the scope of the font declaration.
% There is no need to engage in any other typographic manipulation.

\bibliography{mnrasmnemonic,refs}
\bibliographystyle{mnrasv2}
\end{document}